\documentclass{article}

\usepackage{amsmath,stmaryrd}
\usepackage{amsfonts,amssymb}
\usepackage{pstricks}

\newdimen\proofrulebreadth \proofrulebreadth=.05em
\newdimen\proofdotseparation \proofdotseparation=1.25ex
\newdimen\proofrulebaseline \proofrulebaseline=2ex
\newcount\proofdotnumber \proofdotnumber=3
\let\then\relax
\def\hfi{\hskip0pt plus.0001fil}
\mathchardef\squigto="3A3B
\newif\ifinsideprooftree\insideprooftreefalse
\newif\ifonleftofproofrule\onleftofproofrulefalse
\newif\ifproofdots\proofdotsfalse
\newif\ifdoubleproof\doubleprooffalse
\let\wereinproofbit\relax
\newdimen\shortenproofleft
\newdimen\shortenproofright
\newdimen\proofbelowshift
\newbox\proofabove
\newbox\proofbelow
\newbox\proofrulename
\def\shiftproofbelow{\let\next\relax\afterassignment\setshiftproofbelow\dimen0 }
\def\shiftproofbelowneg{\def\next{\multiply\dimen0 by-1 }%
\afterassignment\setshiftproofbelow\dimen0 }
\def\setshiftproofbelow{\next\proofbelowshift=\dimen0 }
\def\setproofrulebreadth{\proofrulebreadth}

\def\prooftree{
\ifnum  \lastpenalty=1
\then   \unpenalty
\else   \onleftofproofrulefalse
\fi
\ifonleftofproofrule
\else   \ifinsideprooftree
        \then   \hskip.5em plus1fil
        \fi
\fi
\bgroup
\setbox\proofbelow=\hbox{}\setbox\proofrulename=\hbox{}%
\let\justifies\proofover\let\leadsto\proofoverdots\let\Justifies\proofoverdbl
\let\using\proofusing\let\[\prooftree
\ifinsideprooftree\let\]\endprooftree\fi
\proofdotsfalse\doubleprooffalse
\let\thickness\setproofrulebreadth
\let\shiftright\shiftproofbelow \let\shift\shiftproofbelow
\let\shiftleft\shiftproofbelowneg
\let\ifwasinsideprooftree\ifinsideprooftree
\insideprooftreetrue
\setbox\proofabove=\hbox\bgroup$\displaystyle 
\let\wereinproofbit\prooftree
\shortenproofleft=0pt \shortenproofright=0pt \proofbelowshift=0pt
\onleftofproofruletrue\penalty1
}

\def\eproofbit{
\ifx    \wereinproofbit\prooftree
\then   \ifcase \lastpenalty
        \then   \shortenproofright=0pt  
        \or     \unpenalty\hfil         
        \or     \unpenalty\unskip       
        \else   \shortenproofright=0pt  
        \fi
\fi
\global\dimen0=\shortenproofleft
\global\dimen1=\shortenproofright
\global\dimen2=\proofrulebreadth
\global\dimen3=\proofbelowshift
\global\dimen4=\proofdotseparation
\global\count255=\proofdotnumber
$\egroup  
\shortenproofleft=\dimen0
\shortenproofright=\dimen1
\proofrulebreadth=\dimen2
\proofbelowshift=\dimen3
\proofdotseparation=\dimen4
\proofdotnumber=\count255
}

\def\proofover{
\eproofbit 
\setbox\proofbelow=\hbox\bgroup 
\let\wereinproofbit\proofover
$\displaystyle
}%
\def\proofoverdbl{
\eproofbit 
\doubleprooftrue
\setbox\proofbelow=\hbox\bgroup 
\let\wereinproofbit\proofoverdbl
$\displaystyle
}%
\def\proofoverdots{
\eproofbit 
\proofdotstrue
\setbox\proofbelow=\hbox\bgroup 
\let\wereinproofbit\proofoverdots
$\displaystyle
}%
\def\proofusing{
\eproofbit 
\setbox\proofrulename=\hbox\bgroup 
\let\wereinproofbit\proofusing
\kern0.3em$
}

\def\endprooftree{
\eproofbit 
  \dimen5 =0pt
\dimen0=\wd\proofabove \advance\dimen0-\shortenproofleft
\advance\dimen0-\shortenproofright
\dimen1=.5\dimen0 \advance\dimen1-.5\wd\proofbelow
\dimen4=\dimen1
\advance\dimen1\proofbelowshift \advance\dimen4-\proofbelowshift
\ifdim  \dimen1<0pt
\then   \advance\shortenproofleft\dimen1
        \advance\dimen0-\dimen1
        \dimen1=0pt
        \ifdim  \shortenproofleft<0pt
        \then   \setbox\proofabove=\hbox{%
                        \kern-\shortenproofleft\unhbox\proofabove}%
                \shortenproofleft=0pt
        \fi
\fi
\ifdim  \dimen4<0pt
\then   \advance\shortenproofright\dimen4
        \advance\dimen0-\dimen4
        \dimen4=0pt
\fi
\ifdim  \shortenproofright<\wd\proofrulename
\then   \shortenproofright=\wd\proofrulename
\fi
\dimen2=\shortenproofleft \advance\dimen2 by\dimen1
\dimen3=\shortenproofright\advance\dimen3 by\dimen4
\ifproofdots
\then
        \dimen6=\shortenproofleft \advance\dimen6 .5\dimen0
        \setbox1=\vbox to\proofdotseparation{\vss\hbox{$\cdot$}\vss}%
        \setbox0=\hbox{%
                \advance\dimen6-.5\wd1
                \kern\dimen6
                $\vcenter to\proofdotnumber\proofdotseparation
                        {\leaders\box1\vfill}$%
                \unhbox\proofrulename}%
\else   \dimen6=\fontdimen22\the\textfont2 
        \dimen7=\dimen6
        \advance\dimen6by.5\proofrulebreadth
        \advance\dimen7by-.5\proofrulebreadth
        \setbox0=\hbox{%
                \kern\shortenproofleft
                \ifdoubleproof
                \then   \hbox to\dimen0{%
                        $\mathsurround0pt\mathord=\mkern-6mu%
                        \cleaders\hbox{$\mkern-2mu=\mkern-2mu$}\hfill
                        \mkern-6mu\mathord=$}%
                \else   \vrule height\dimen6 depth-\dimen7 width\dimen0
                \fi
                \unhbox\proofrulename}%
        \ht0=\dimen6 \dp0=-\dimen7
\fi
\let\doll\relax
\ifwasinsideprooftree
\then   \let\VBOX\vbox
\else   \ifmmode\else$\let\doll=$\fi
        \let\VBOX\vcenter
\fi
\VBOX   {\baselineskip\proofrulebaseline \lineskip.2ex
        \expandafter\lineskiplimit\ifproofdots0ex\else-0.6ex\fi
        \hbox   spread\dimen5   {\hfi\unhbox\proofabove\hfi}%
        \hbox{\box0}%
        \hbox   {\kern\dimen2 \box\proofbelow}}\doll%
\global\dimen2=\dimen2
\global\dimen3=\dimen3
\egroup 
\ifonleftofproofrule
\then   \shortenproofleft=\dimen2
\fi
\shortenproofright=\dimen3
\onleftofproofrulefalse
\ifinsideprooftree
\then   \hskip.5em plus 1fil \penalty2
\fi
}

\usepackage{xy}
\xyoption{all}
 \xyoption{line}
\CompileMatrices
\usepackage{graphicx}

\newcommand{\Set}{{\sf Set}}
\newcommand{\Pfn}{{\sf Pfn}}
\newcommand{\Vect}{{\sf Vec}}

\newcommand{\Rel}{{\sf Rel}}

\newcommand{\id}{{\rm id}}

\newcommand{\CCC}{{\cal C}}
\newcommand{\DDD}{{\cal D}}

\renewcommand{\Bbb}{\mathbb}

\newcommand{\BBb}{{\Bbb B}}
\newcommand{\CCc}{{\Bbb C}}

\newcommand{\NNn}{{\Bbb N}}

\newcommand{\TTt}{{\Bbb T}}

\newcounter{countroman}
{\begin{list}{{\rm (\roman{countroman})}}
{\usecounter{countroman}}}%
{\end{list}}

\newcounter{countbarabic}
{\begin{list}{{\rm (\arabic{countbarabic})}}
{\usecounter{countbarabic}}}%
{\end{list}}

\newcounter{countalpha}
{\begin{list}{(\alph{countalpha})}{\usecounter{countalpha}}}%
{\end{list}}

\newcounter{countalphabf}
{\protect\begin{list}{{\rm (}{\bf \protect\alph{countalphabf}}{\rm
)}}{\protect\usecounter{countalphabf}}}%
{\end{list}}

\newcommand{\bitmz}{\vspace{-.5\baselineskip}
\begin{itemize}}
\newcommand{\eitmz}{\end{itemize}\vspace{-.
25\baselineskip}}

\newcommand{\bdesc}{\vspace{-.5\baselineskip}
\begin{description}}
\newcommand{\edesc}{\end{description}\vspace{-.
25\baselineskip}}

\mathcode`\<="4268 
\mathcode`\>="5269 
\mathchardef\gt="313E 
\mathchardef\lt="313C 
\newsavebox{\barr}
\savebox{\barr}{\hspace*{-9.5pt}\raisebox{1.25pt}{$
\scriptscriptstyle%
|$}\hspace*{4.5pt}} 
\newsavebox{\barrleft}
\savebox{\barrleft}{\hspace*{-8.5pt}\raisebox{1.25pt}{$
\scriptscriptstyle%
|$}\hspace*{10pt}}

 \def\pushright#1{{
    \parfillskip=0pt            
    \widowpenalty=10000         
    \displaywidowpenalty=10000  
    \finalhyphendemerits=0      
    \leavevmode                 
    \unskip                     
    \nobreak                    
    \hfil                       
    \penalty50                  
    \hskip.2em                  
    \null                       
    \hfill                      
    {#1}                        
    \par}}                      

 \def\qed{\pushright{$\square$}\penalty-700 \smallskip}

\newenvironment{prf}[1]{\begin{trivlist} \item[{\bf ~Proof}#1.]}%
{\qed\end{trivlist}}

\newcommand{\be}[1]{\begin{#1}}

\newcommand{\ee}[1]{\end{#1}}
\newcommand{\beq}{\begin{equation}}
\newcommand{\eeq}{\end{equation}}
\newcommand{\ba}[1]{\begin{array}{#1}}
\newcommand{\ea}{\end{array}}
\newcommand{\bea}{\begin{eqnarray}}
\newcommand{\eea}{\end{eqnarray}}
\newcommand{\bear}{\begin{eqnarray*}}
\newcommand{\eear}{\end{eqnarray*}}
\newcommand{\bpr}{\begin{prf}{}}
\newcommand{\epr}{\end{prf}}
\newcommand{\bprf}[1]{\begin{prf}{#1}}
\newcommand{\eprf}{\end{prf}}

\newtheorem{thm}{Theorem}[section]

\newtheorem{prop}[thm]{Proposition}

\newtheorem{corollary}[thm]{Corollary}
\newtheorem{cond}{}[thm]

\newtheorem{prenumb}[thm]{\hspace{-1ex}}

\renewcommand{\to}{\rightarrow}
\newcommand{\ot}{\leftarrow}

\newcommand{\tto}[1]{\xrightarrow{#1}}

\newcommand{\oot}[1]{\xleftarrow{#1}}

\newcommand{\epi}{\twoheadrightarrow}

\newcommand{\eepi}[1]{\stackrel{#1}{\twoheadrightarrow}}

\newcommand{\pfn}{\rightharpoonup}

\newcommand{\undefined}{\uparrow}

\newcommand{\Source}{\XXX}

\newcommand{\supp}[1]{{\rm supp}\left({#1}\right)}

\newcommand{\comp}{\, ;}
\newcommand{\pcomp}{\|}

\newcommand{\Nn}{\Xi}

\newcommand{\UK}{u}
\newcommand{\SK}{s}
\newcommand{\TIME}{tm}
\newcommand{\SPACE}{sp}
\newcommand{\CX}{cx}

\newcommand{\cata}[1]{\llparenthesis {#1} \rrparenthesis}

\newcommand{\leqad}{\stackrel {\scriptstyle +} \leq}

\newcommand{\eqad}{\stackrel {\scriptstyle +} =}

\newcommand{\enco}[1]{\left\ulcorner{#1}\right\urcorner}
\newcommand{\deco}[1]{\left\{{#1}\right\}}

\newcommand{\mnd}{\varrho}
\newcommand{\unt}{{\scriptstyle \bot}}
\newcommand{\cmn}{\delta}
\newcommand{\cun}{{\scriptstyle \top}}

\newcommand{\doublone}{1\!\!1}
\newcommand{\tru}{t\!t}
\newcommand{\fls}{f\!\!f}
\newcommand{\Bits}{\BBb}
\newcommand{\Totel}{\TTt}
\newcommand{\Nnn}{\widehat\Nn}
\newcommand{\ifthenelse}{{\sf ifthenelse}}
\newcommand{\Base}[1]{{#1}^\flat}

\newcommand{\diag}{\lozenge}
\newcommand{\codiag}{\lozenge^o}
\newcommand{\bang}{!}

\newcommand{\RC}[1]{{#1}_{\DDD}}

\title{Monoidal computer I:\\ 
Basic computability by string diagrams}
\author{Dusko Pavlovic\\
Royal Holloway and Twente\\
dusko.pavlovic@rhul.ac.uk
}

\date{}
\begin{document} 
\maketitle

\begin{abstract}
We present  a new model of computation, described in terms of monoidal categories. It conforms the Church-Turing Thesis, and captures the same computable functions as the standard models. It provides a succinct categorical interface to most of them, free of their diverse implementation details, using the ideas and structures that in the meantime emerged from research in semantics of computation and programming. The salient feature of the language of monoidal categories is that it is supported by a sound and complete graphical formalism, string diagrams, which provide a concrete and intuitive interface for abstract reasoning about computation. The original motivation and the ultimate goal of this effort is to provide a convenient high level programming language for a theory of computational resources, such as one-way functions, and trapdoor functions, by adopting the methods for hiding the low level implementation details that emerged from practice. In the present paper, we make a first step towards this ambitious goal, and sketch a path to reach it. This path is pursued in three sequel papers, that are in preparation. 
\end{abstract}

\section{Introduction}
\subsection{Preamble: Resources as one-way functions}
\begin{figure}
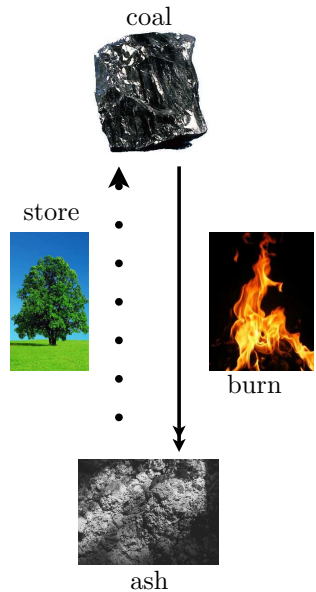

\centering
\newcommand{\Resource}{\includegraphics[height=2cm]{PIC/coal.epsf}}
\newcommand{\Residue}{\includegraphics[height=1.4cm]{PIC/ashes.epsf}}
\newcommand{\utility}{\includegraphics[height=1.85cm]{PIC/fire-flames.epsf}}
\newcommand{\investment}{\includegraphics[height=1.85cm]{PIC/tree.epsf}}
\newcommand{\coal}{coal}
\newcommand{\ashes}{ash}
\newcommand{\burn}{burn}
\newcommand{\store}{store}
\def\JPicScale{.8}
\input{PIC/resource-photos.tex}
\caption{A resource is easy to use, but slow to accumulate}
\label{fig-resource-coal}
\end{figure}
\begin{figure}
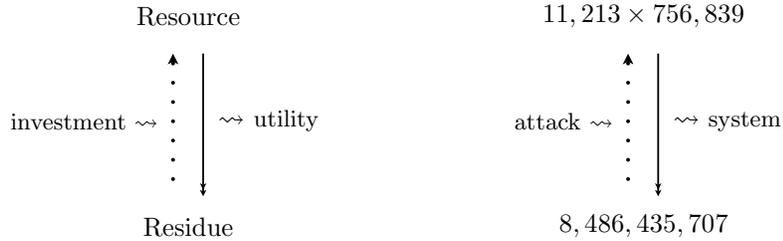

\begin{minipage}[b]{0.43\linewidth}
\centering
\newcommand{\Resource}{Resource}
\newcommand{\Residue}{Residue}
\newcommand{\utility}{$\rightsquigarrow$ \small utility}
\newcommand{\investment}{\small investment $\rightsquigarrow$}
\def\JPicScale{.4}
\input{PIC/resource.tex}
\end{minipage}
\hspace{.6cm}
\begin{minipage}[b]{0.43\linewidth}
\centering
\newcommand{\Resource}{$11,213 \times 756,839$}
\newcommand{\Residue}{$8,486,435,707$}
\newcommand{\utility}{$\rightsquigarrow$ \small system}
\newcommand{\investment}{\small attack $\rightsquigarrow$}
\def\JPicScale{.4}
\input{PIC/resource.tex}
\end{minipage}
\caption{Resources support one-way functions}
\label{fig-resource}
\end{figure}

What is a resource? A typical example of a resource is coal: we burn it to get heat. But while the process of burning coal is easy and relatively quick, the processes of capturing the energy in plants and of fossilizing them into coal, take millions of years. This asymmetry can be seen as the characteristic property of resources: they are easy to use, but hard to come by. The difference between the utility of consuming a resource and the investment needed to produce it is what makes it into a resource, as illustrated on Fig.\ref{fig-resource}. 


Large parts of modern cryptography are based on the assumptions that some easy computational operations are hard to invert: e.g., that the exponents in finite fields are much easier to compute than logarithms, and that multiplying integers is in many cases substantially easier than factoring them. Pairs of large primes are thus used as security resources, allowing the system to easily hide a secret prime by multiplying it with another secret prime, and leaving the attacker with the hard task of factoring the product. In a sense, the resources can thus be construed as one-way functions. The theory of one-way functions, that underlies modern cryptography, can thus be viewed as a computational formalization of the basic idea of a resource.

But it seems remarkable that such a simple idea requires such a delicate formalization. The theory of one-way functions has so far not even proved that one-way functions exist! This might be a temporary state of affairs; but a proof that one-way functions do exist would yield a proof of the great $P\neq NP$ conjecture, which is not thought to be within reach at the moment. Moreover, this is not the only shaky point of the theory. E.g., we have also not proven that the existence of one-way functions would imply the existence of trapdoor functions, which also seem necessary for a practical cryptography, as they are the cryptographic locks, that allow those with the key in, and leave those without the key out. There is indeed a whole hierarchy of unproven hypotheses about the exploitability of computational hardness as a security resource  \cite{ImpagliazzoR:universes}.

So why is the notion of a computational resource, \emph{viz.}\/ of a one-way function, so brittle? On one hand, this is a deep question, that cuts into the tissue of modern mathematics \cite{LevinL:tale}, with the philosophical implications beyond the scope of our technical analyses. On the other hand, the emerging problems of cyber security seem to lead beyond one-way functions over data, to one-way program transformations \cite{PavlovicD:NSPW11}, and bring this theoretic question into the realm of everyday security practices.

\paragraph{Outline of the program.} In the present paper, we begin a journey towards a high level language for computational resources, or one-way functions, interpreted broadly. At this first step, we introduce the main vehicle, monoidal computer, and spell out the elements of computability in it. The only claim is that it is a convenient conduit, providing some insightful pictures of some subtle concepts. In the second issue of the planned series of papers, we shall study abstract complexity measures in monoidal computer, derive the time complexity and the space complexity measures as natural special cases, and spell out the elements of complexity theory. The third step will lead into randomized computation which will naturally, and perhaps not entirely unexpectedly, be captured through some familiar constructions of categorical algebra. The standard notions of one-way function and of trapdoor function will come within reach in that part. In the fourth part we are hoping to use monoidal computer to capture some parts of algorithmic information theory and pursue the idea of one-way algorithm transformation, as a logical resource of security, proposed in \cite{PavlovicD:NSPW11}. Each part seems to be more interesting, and more challenging than the previous one. At the moment, the second and the third parts exist as fairly detailed working papers, with most of the proofs, whereas the fourth part is still a handwritten sketch.

\subsection{The idea of a monoidal computer}
We have been programming computers for almost 70 years. This extensive practice has engendered a large variety of programming languages, enabling us in many domains to convey to computers our high level views of our algorithmic ideas, while allowing us to leave the implementation details for later, or some of them even to the computers themselves. The abstraction tools are the crucial components of programmer's toolkit, continuously spreading through an ever wider range of programming, scripting and specification languages.

But while the programming practices are substantially facilitated by the high-level languages, and by the evolved methods of abstraction, the research in computability, complexity and cryptography still involves a great amount of low level programming. The Church-Turing Thesis asserts that the various models of computation have equivalent computational powers, and thus capture the same notion of computability; yet the sheer variety of these models shows that each of them contains irrelevant implementation details. Writing $\lambda$-expressions and designing Turing machines are often pleasant as mathematical exercises, but they painfully resemble machine programming when it comes, e.g., to proving security of a crypto system. Security proofs therefore often require an enormous amount of effort to write, and sometimes an even greater amount of effort to read. Several solutions have been proposed. One family of solutions, pursued with a great success in the formal methods community, is based on automated evaluation of $\lambda$-expressions \cite{BartheG:CRYPTO11}. Another family of solutions, endorsed by a majority of working cryptographers, and thus undoubtedly very successful as well, is to present algorithms in one of the various versions of pseudo-code, referring to a tacit Turing machine formalism, which is accepted to be too verbose and too routinely to be fully spelled out in research papers. 

Mostly as a thought experiment, we contemplate yet another kind of a solution. If machine programming has been encapsulated into the high level programming languages in the practice of computation, maybe the same can be done for the theory. So let us try to specify computer as a virtual function, or as an abstract data type: a mere interface for reader's favorite model of computation.


\paragraph{Outline of the paper.} In Sec.~\ref{Sec-cats} we provide a brief overview of the basic categorical concepts to be used. In Sec.~\ref{Sec-data-services} we introduce the categorical structure that provides in monoidal computers the data services, such as copying, deleting and filtering. Sec.~\ref{Sec-Moncom} presents the formal definition and the basic examples of monoidal computer. A method to implement in monoidal computer the basic logical and arithmetic constructions is proposed in Sec.~\ref{Sec-Arithmetic}. A basic fixed point construction is drawn in Sec.~\ref{Sec-Fixpoints}, and extended in Sec.~\ref{Sec-Kleene} into a diagrammatic proof of Kleene's Second Recursion Theorem. Sec.~\ref{Sec-Halting} derives a similar proof that the Halting Problem is undecidable, and Sec.~\ref{Sec-Rice} completes the paper by Rice's Theorem, which says that every nontrivial predicate over computations must be undecidable. In the final section we discuss the ideas that will be pursued in the sequel.

\section{Monoidal categories and string diagrams}\label{Sec-cats}
We begin with an informal overview of the monoidal categories, albeit of the small fragment of the structure that will be used in this paper. More thorough introductions can be found in many basic texts on categories, e.g. \cite{MacLaneS:CWM,KellyM:Enriched,Joyal-Street:geometry}.

\begin{figure}
\newcommand{\ah}{A}
\newcommand{\bh}{B}
\newcommand{\ch}{C}
\renewcommand{\dh}{D}
\newcommand{\ih}{I}
\newcommand{\aah}{}
\newcommand{\chh}{}
\newcommand{\af}{f}
\newcommand{\ag}{g}
\newcommand{\ahh}{h}
\newcommand{\ak}{k}
\def\JPicScale{.9}
\centering
\renewcommand{\aah}{A}\renewcommand{\chh}{C}
\input{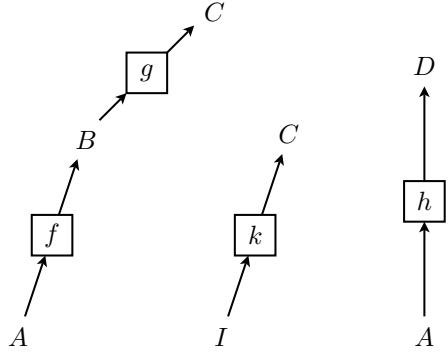}
\caption{String diagrams}
\label{fig-String-diagrams}
\end{figure}

A monoidal category $\CCC$ as a universe of 
\begin{itemize}
\item objects (or data types) $A, B,\ldots, L, M\ldots \in |\CCC|$ and 
\item morphisms (or computations) $f, g\ldots \in \CCC(A,B), u, t\ldots \in \CCC(X, Y)$.
\end{itemize}
The morphisms are often also written in the form $f:A\to B$, or  $A\tto f B$, and in the form of \emph{string diagrams}, as on Fig.~\ref{fig-String-diagrams}
The categorical structure captures
\begin{itemize}
\item sequential composition 
\bear
\CCC(A,B) \times \CCC(B,C) & \to & \CCC(A,C)\\
<f,g> &\longmapsto & g\circ f
\eear
also written $A\tto f B \tto g C$,
\item parallel composition 
\bear
\CCC(A,B) \times \CCC(L,M) & \to & \CCC(A\otimes L,C\otimes M)\\
<f,t> &\longmapsto & f\otimes t
\eear
also written $A\otimes L \tto{f\otimes t} B\otimes M$.
\end{itemize}
The units for the above structures are
\begin{itemize}
\item the \emph{identities} $\id_A:A\to A$, satisfying
\beq f\circ \id_A\ = \ f\ =\ \id_B\circ f\eeq
\item the \emph{unit type} $I\in |\CCC|$, satisfying
\beq\label{eq-unitary}X \otimes I\ =\ X\ =\ I\otimes X\eeq
\end{itemize}

\begin{figure}
\newcommand{\bh}{B}
\newcommand{\ch}{C}
\renewcommand{\dh}{D}
\newcommand{\ih}{I}
\newcommand{\aah}{A}
\newcommand{\chh}{C}
\newcommand{\af}{f}
\newcommand{\ag}{g}
\newcommand{\ahh}{h}
\newcommand{\ak}{k}
\newcommand{\lhs}{(g\otimes C\otimes D)\circ (f\otimes C\otimes h)\circ (A\otimes k\otimes A)} 
\newcommand{\eqls}{}
\newcommand{\eqlsmall}{=}
\newcommand{\simmetry}{\scriptstyle \varsigma}
\newcommand{\lhsup}{(g\circ f)\otimes k \otimes h}
\newcommand{\lhsdown}{(g\otimes C\otimes D)\circ (B\otimes k\otimes D)\circ (f\otimes h)}
\newcommand{\rhsup}{(g\otimes C\otimes h)\circ ((\varsigma \circ (k\otimes f))\otimes A)}
\def\JPicScale{.9}
\centering
\input{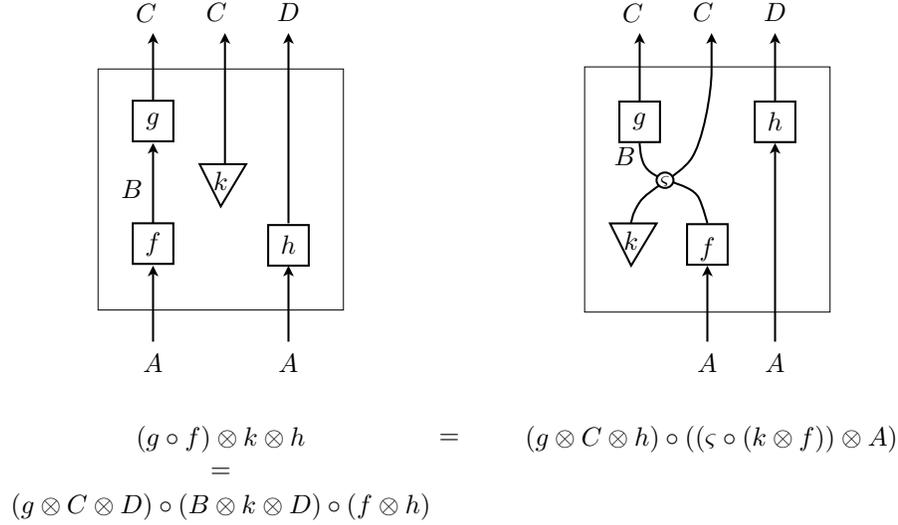}
\caption{String diagrams are sound and complete for monoidal equations}
\label{fig-String-diagram-equations}
\end{figure}

\paragraph{Elements.} An "element" $k$ of type $C$ in $\CCC$ is viewed as a morphism $I\tto k C$, where $I$ is the tensor unit. In the diagrams, $I$ and its strings are usually elided, and the elements are drawn as the triangles pointing down, like in Fig.~\ref{fig-String-diagram-equations}. Dually, the morphisms into $I$ are drawn as triangles pointing up, like in Fig.~\ref{fig-data-service}. The set of elements $\CCC(I,A)$ is often abbreviated to $\CCC(A)$.

\paragraph{Scalars.} The elements of the monoidal unit $I$ are usually called (abstract) \emph{scalars}\/ the monoidal category $\CCC$. The set of scalars is thus $\CCC(I) = \CCC(I,I)$.

\paragraph{Examples.} A typical example of a monoidal category is the category $\Vect$ of vector spaces and linear operators  over a ground field  $I$. The monoidal structure is provided by the standard tensor product $\otimes$, and the field $I$ is the unit with respect to it. The elements of $I$ are, of course, the scalars in the usual sense. They are captured as the abstract scalars, i.e. the linear operators $\Vect(I) = \Vect(I,I)$, because each scalar $r\in I$ determines a unique linear operator $I\tto{r\cdot (-)} I$. The categories $\Set$ of sets and functions, and $\Rel$ of sets and relations provide further examples of monoidal categories. In fact, each of them is a monoidal category both with \begin{itemize}
\item the multiplicative structure $(\times, \doublone)$, where $A\times B$ is the cartesian product of the sets $A$ and $B$, and $\doublone$ is a one element set; and also with 
\item the additive structure $(+, \emptyset)$, where $A+B$ is the disjoint union of the sets $A$ and $B$, and $\emptyset$ is the empty set. 
\end{itemize}
Note that $\Set(\doublone)$ and $\Rel(\emptyset)$ each contain a unique scalar, whereas $\Rel(\doublone)$ has two scalars. Any monoid can also be viewed as a discrete monoidal category, with the identities as the only morphisms, and thus with single scalar.

The categories $\Set_{fin}$ and $\Rel_{fin}$ of \emph{finite\/} sets and all functions, resp. all relations between those sets, are equivalent with their \emph{skeletal}\/ subcategories spanned by the natural numbers, viewed as finite sets. The upshot of this reduction is that the monoidal structures can now be defined to be \emph{strict}: e.g., while the cartesian products $(A\times B)\times C$ and $A\times (B\times C)$ are isomorphic along a pair of bijective functions, uniformly defined for all sets $A, B, C$, if we restrict to natural numbers $a,b,c \in \NNn$, then the products $(a\times b)\times c$ and $a\times (b\times c)$ denote the same number. The isomorphism $(A\times B)\times C\cong A\times (B\times C)$ and $A\times \doublone \cong A \cong \doublone \times A$ can be strengthened to \emph{strict}\/ equalities $(a\times b)\times c = a\times (b\times c)$ and $a\times 1 = a = 1\times a$. The same holds for the additive structure. The same holds for the additive structure.

\paragraph{Assumptions.} In the present paper, the monoidal structure is always assumed to be
\begin{itemize}
\item \emph{strict}, meaning that the tensor is strictly unitary, and strictly associative, in the sense that they satisfy
\bea
A\otimes I\ =\ & A & \ =\ I\otimes A\label{eq-I}\\
(A\otimes B) \otimes C & = & A\otimes (B\otimes C)\label{eq-assoc}
\eea
so that we can drop the brackets; and
\item \emph{symmetric}, in the sense that is a family of isomorphisms
\bea \label{eq-symmm}
A\otimes B & \stackrel \varsigma\cong & B\otimes A
\eea
indexed by $A,B\in |\CCC|$, satisfying the standard coherence requirements \cite{MacLaneS:CWM,KellyM:Enriched}.
\end{itemize}
The symmetries $\varsigma$ in \eqref{eq-symmm} must be kept as explicit isomorphisms; strictifying them like (\ref{eq-I}-\ref{eq-assoc}) would lead to degenerate categories.

\paragraph{Graphic notation.} The string diagrams for monoidal categories were formally developed in \cite{Joyal-Street:geometry}, but  go back at least to \cite{Penrose}. In a formal sense, their geometric transformations capture precisely the algebraic laws of monoidal categories, or of the parallel and the sequential compositions of computations. The string intersections correspond to the tensor symmetries. Fig.~\ref{fig-String-diagram-equations} illustrates the correspondence of the monoidal equations and string diagrams.

\section{Data services}\label{Sec-data-services}
Computation requires moving the data around, copying them for reuse, deleting what is not needed, comparing the various values. These operations form an interesting algebraic structure, consisting of a \emph{comonoid} and a  \emph{semigroup}, conjoined together by a coherence requirement known in algebra as the \emph{Frobenius condition}. In this section we spell out these algebraic structures in the monoidal framework.

\subsection{Basic structures}
\subsubsection{Monoids and semigroups}
A \emph{semigroup\/} is usually defined as a set with an associative binary operation. In a monoidal category, the associativity of a binary operation $A\otimes A\stackrel{\mnd}\to A$ means that it satisfies the equation
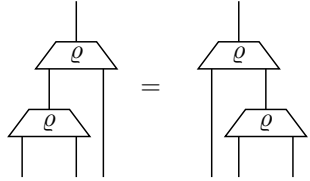
\beq
\mnd \circ (\mnd \otimes A)\  =\   \mnd \circ (A\otimes \mnd)
\eeq
\begin{center}
\newcommand{\comonoidd}{\mnd}
\newcommand{\comontwo}{\mnd}
\def\JPicScale{.9}
\ifx\JPicScale\undefined\def\JPicScale{1}\fi
\psset{unit=\JPicScale mm}
\psset{linewidth=0.3,dotsep=1,hatchwidth=0.3,hatchsep=1.5,shadowsize=1,dimen=middle}
\psset{dotsize=0.7 2.5,dotscale=1 1,fillcolor=black}
\psset{arrowsize=1 2,arrowlength=1,arrowinset=0.25,tbarsize=0.7 5,bracketlength=0.15,rbracketlength=0.15}
\begin{pspicture}(0,0)(45,26)
\psline[linewidth=0.22](15,16)(15,-0)
\pspolygon[linewidth=0.22](17,16)
(5,16)
(8,20)
(14,20)(17,16)
\psline[linewidth=0.22](11,20)(11,26)
\psline[linewidth=0.22](7,16)(7,10)
\psline[linewidth=0.22](11,6)(11,-0)
\pspolygon[linewidth=0.22](13,6)
(1,6)
(4,10)
(10,10)(13,6)
\psline[linewidth=0.22](3,6)(3,-0)
\rput(22,13){$=$}
\psline[linewidth=0.22](31,16)(31,-0)
\pspolygon[linewidth=0.22](41,16)
(29,16)
(32,20)
(38,20)(41,16)
\psline[linewidth=0.22](35,20)(35,26)
\psline[linewidth=0.22](39,16)(39,10)
\psline[linewidth=0.22](43,6)(43,-0)
\pspolygon[linewidth=0.22](45,6)
(33,6)
(36,10)
(42,10)(45,6)
\psline[linewidth=0.22](35,6)(35,-0)
\rput(39,8){$\comonoidd$}
\rput(35,18){$\comonoidd$}
\rput(7,8){$\comonoidd$}
\rput(11,18){$\comonoidd$}
\end{pspicture}

\end{center}
A monoid is, of course, a semigroup with a unit. The structure of a comonoid is dual to that of monoid, i.e. a pair of arrows  $A\otimes A\stackrel{\cmn}\ot A \stackrel{\cun}{\to}I$ satisfying the equations
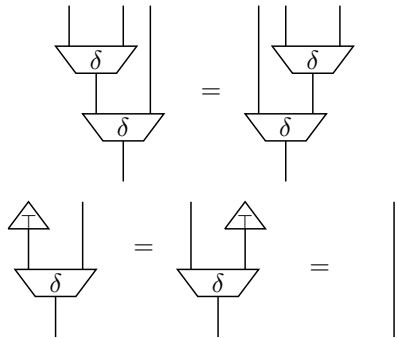
\begin{gather}
(\cmn \otimes A) \circ \cmn  =  (A\otimes \cmn)\circ \cmn\\
(\cun \otimes A) \circ \cmn  = (A\otimes \cun) \circ \cmn = \id_A
\end{gather}
\begin{center}
\newcommand{\comonoidd}{\cmn}
\newcommand{\comonunn}{\cun}
\newcommand{\comontwo}{\cmn}
\def\JPicScale{.9}
\ifx\JPicScale\undefined\def\JPicScale{1}\fi
\psset{unit=\JPicScale mm}
\psset{linewidth=0.3,dotsep=1,hatchwidth=0.3,hatchsep=1.5,shadowsize=1,dimen=middle}
\psset{dotsize=0.7 2.5,dotscale=1 1,fillcolor=black}
\psset{arrowsize=1 2,arrowlength=1,arrowinset=0.25,tbarsize=0.7 5,bracketlength=0.15,rbracketlength=0.15}
\begin{pspicture}(0,0)(60,47.5)
\rput(40.75,6.5){}
\psline[linewidth=0.22](40,31.5)(40,47.5)
\pspolygon[linewidth=0.22](38,31.5)
(50,31.5)
(47,27.5)
(41,27.5)(38,31.5)
\psline[linewidth=0.22](44,27.5)(44,21.5)
\psline[linewidth=0.22](6,8.5)(6,14.5)
\psline[linewidth=0.22](3,14.5)
(6,18.5)
(9,14.5)(3,14.5)
\psline[linewidth=0.22](48,31.5)(48,37.5)
\psline[linewidth=0.22](44,41.5)(44,47.5)
\pspolygon[linewidth=0.22](42,41.5)
(54,41.5)
(51,37.5)
(45,37.5)(42,41.5)
\psline[linewidth=0.22](52,41.5)(52,47.5)
\rput(33,34.5){$=$}
\psline[linewidth=0.22](24,31.5)(24,47.5)
\pspolygon[linewidth=0.22](14,31.5)
(26,31.5)
(23,27.5)
(17,27.5)(14,31.5)
\psline[linewidth=0.22](20,27.5)(20,21.5)
\psline[linewidth=0.22](16,31.5)(16,37.5)
\psline[linewidth=0.22](12,41.5)(12,47.5)
\pspolygon[linewidth=0.22](10,41.5)
(22,41.5)
(19,37.5)
(13,37.5)(10,41.5)
\psline[linewidth=0.22](20,41.5)(20,47.5)
\psline[linewidth=0.22](30,8.5)(30,18.5)
\pspolygon[linewidth=0.22](28,8.5)
(40,8.5)
(37,4.5)
(31,4.5)(28,8.5)
\psline[linewidth=0.22](34,4.5)(34,-1.5)
\rput(23,11.5){$=$}
\psline[linewidth=0.22](14,8.5)(14,18.5)
\pspolygon[linewidth=0.22](4,8.5)
(16,8.5)
(13,4.5)
(7,4.5)(4,8.5)
\psline[linewidth=0.22](10,4.5)(10,-1.5)
\psline[linewidth=0.22](60,-1.5)(60,18.5)
\rput(49,8.5){$=$}
\psline[linewidth=0.22](38,8.5)(38,14.5)
\psline[linewidth=0.22](35,14.5)
(38,18.5)
(41,14.5)(35,14.5)
\rput(16,39.5){$\comonoidd$}
\rput(20,29.5){$\comonoidd$}
\rput(48,39.5){$\comonoidd$}
\rput(44,29.5){$\comonoidd$}
\rput(34,6.5){$\comonoidd$}
\rput(10,6.5){$\comonoidd$}
\rput(6,15.5){$\comonunn$}
\rput(38,15.5){$\comonunn$}
\end{pspicture}

\end{center}
Both semigroups and comonoids are said to be \emph{commutative}\/ when they remain unchanged under the composition with the symmetry $A\otimes A \stackrel \varsigma \cong A\otimes A$, as shown on the next diagram.
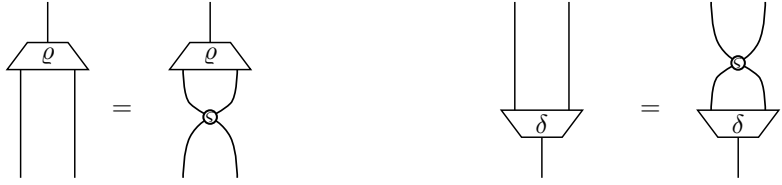
\begin{center}
\newcommand{\comonoidd}{\mnd}
\newcommand{\monoidd}{\cmn}
\newcommand{\simmetry}{\scriptstyle \varsigma}
\def\JPicScale{.9}
\ifx\JPicScale\undefined\def\JPicScale{1}\fi
\psset{unit=\JPicScale mm}
\psset{linewidth=0.3,dotsep=1,hatchwidth=0.3,hatchsep=1.5,shadowsize=1,dimen=middle}
\psset{dotsize=0.7 2.5,dotscale=1 1,fillcolor=black}
\psset{arrowsize=1 2,arrowlength=1,arrowinset=0.25,tbarsize=0.7 5,bracketlength=0.15,rbracketlength=0.15}
\begin{pspicture}(0,0)(119,26)
\psline[linewidth=0.22](15,16)(15,-0)
\pspolygon[linewidth=0.22](17,16)
(5,16)
(8,20)
(14,20)(17,16)
\psline[linewidth=0.22](11,20)(11,26)
\rput(22,10){$=$}
\pspolygon[linewidth=0.22](41,16)
(29,16)
(32,20)
(38,20)(41,16)
\psline[linewidth=0.22](35,20)(35,26)
\rput(35,18){$\comonoidd$}
\rput(11,18){$\comonoidd$}
\psline[linewidth=0.22](7,16)(7,0)
\pscustom[linewidth=0.25]{\psbezier(31,16)(31,16)(31,12)(32,11)
\psbezier(33,10)(35,9)(35,9)
\psbezier(35,9)(37,8)(38,6)
\psbezier(39,4)(39,0)(39,0)
}
\pscustom[linewidth=0.25]{\psbezier(39,16)(39,16)(39,12)(38,11)
\psbezier(37,10)(35,9)(35,9)
\psbezier(35,9)(33,8)(32,6)
\psbezier(31,4)(31,0)(31,0)
}
\rput{0}(35,9){\psellipse[fillcolor=white,fillstyle=solid](0,0)(1,-1)}
\rput(35,9){$\simmetry$}
\psline[linewidth=0.22](80,10)(80,26)
\pspolygon[linewidth=0.22](78,10)
(90,10)
(87,6)
(81,6)(78,10)
\psline[linewidth=0.22](84,6)(84,0)
\rput(100,10){$=$}
\pspolygon[linewidth=0.22](107,10)
(119,10)
(116,6)
(110,6)(107,10)
\psline[linewidth=0.22](113,6)(113,0)
\psline[linewidth=0.22](88,10)(88,26)
\pscustom[linewidth=0.25]{\psbezier(117,10)(117,10)(117,14)(116,15)
\psbezier(115,16)(113,17)(113,17)
\psbezier(113,17)(111,18)(110,20)
\psbezier(109,22)(109,26)(109,26)
}
\pscustom[linewidth=0.25]{\psbezier(109,10)(109,10)(109,14)(110,15)
\psbezier(111,16)(113,17)(113,17)
\psbezier(113,17)(115,18)(116,20)
\psbezier(117,22)(117,26)(117,26)
}
\rput{0}(113,17){\psellipse[fillcolor=white,fillstyle=solid](0,0)(1,-1)}
\rput(113,17){$\simmetry$}
\rput(113,8){$\monoidd$}
\rput(84,8){$\monoidd$}
\end{pspicture}

\end{center}

\subsubsection{Homomorphisms}\label{sec-homomorphisms}
Given comonoids $A\otimes A\oot{\cmn_A} A \tto{\cun_A} I$ and $B\otimes B\oot{\cmn_B} B \tto{\cun_B} B$ in $\CCC$, the morphism $f\in \CCC(A,B)$ is a comonoid homomorphism if 

\beq \label{eq-comon-hom}
\cmn_B \circ f \ =\  (f\otimes f)\circ \cmn_A\qquad\qquad \quad \qquad \cun_B\circ f\ =\ \cun_A
\eeq

\begin{center}
\newcommand{\monoidd}{\cmn}
\newcommand{\fun}{\scriptstyle f}
\newcommand{\One}{A}
\newcommand{\Two}{B}
\newcommand{\delete}{\cun}
\def\JPicScale{1.1}
\ifx\JPicScale\undefined\def\JPicScale{1}\fi
\psset{unit=\JPicScale mm}
\psset{linewidth=0.3,dotsep=1,hatchwidth=0.3,hatchsep=1.5,shadowsize=1,dimen=middle}
\psset{dotsize=0.7 2.5,dotscale=1 1,fillcolor=black}
\psset{arrowsize=1 2,arrowlength=1,arrowinset=0.25,tbarsize=0.7 5,bracketlength=0.15,rbracketlength=0.15}
\begin{pspicture}(0,0)(77.62,21.88)
\rput(17.5,10.62){$=$}
\pspolygon(0.12,16.88)
(12.12,16.88)
(9.38,12.5)
(3.12,12.5)(0.12,16.88)
\rput(6.12,14.88){$\monoidd$}
\pspolygon[](4.38,7.5)(8.12,7.5)(8.12,3.75)(4.38,3.75)
\psline(6.25,3.75)(6.25,0)
\psline(2.5,21.88)(2.5,16.88)
\psline(10,21.88)(10,16.88)
\psline(6.25,12.5)(6.25,7.5)
\rput[r](5,1.25){$\One$}
\rput[r](5,10){$\Two$}
\pspolygon(25.12,8.75)
(37.12,8.75)
(34.38,4.38)
(28.12,4.38)(25.12,8.75)
\rput(31.12,6.75){$\monoidd$}
\pspolygon[](25.62,16.88)(29.38,16.88)(29.38,13.12)(25.62,13.12)
\psline(27.5,21.25)(27.5,16.88)
\psline(27.5,13.12)(27.5,8.75)
\psline(35,13.12)(35,8.75)
\psline(31.25,4.38)(31.25,0)
\rput[r](29.38,0.62){$\One$}
\rput[r](26.88,19.38){$\Two$}
\pspolygon[](33.12,16.88)(36.88,16.88)(36.88,13.12)(33.12,13.12)
\psline(35,21.25)(35,16.88)
\rput(6.25,5.62){$\fun$}
\rput(27.5,15){$\fun$}
\rput(35,15){$\fun$}
\rput[r](26.88,11.25){$\One$}
\rput[l](36.25,11.25){$\One$}
\rput[l](35.62,19.38){$\Two$}
\rput[r](1.25,19.38){$\Two$}
\rput[l](10.62,19.38){$\Two$}
\rput(68.12,10.62){$=$}
\pspolygon[](60,8.75)(63.75,8.75)(63.75,5)(60,5)
\psline(61.88,5)(61.88,0)
\psline(61.88,14.38)(61.88,8.75)
\rput[r](60.62,1.25){$\One$}
\rput[r](60.62,11.88){$\Two$}
\rput(61.88,6.88){$\fun$}
\pscustom[]{\psline(58.88,14.38)(65.12,14.38)
\psline(65.12,14.38)(62,19.38)
\psbezier(62,19.38)(62,19.38)(62,19.38)
\psline(62,19.38)(58.88,14.38)
\closepath}
\rput(61.88,16.38){$\delete$}
\psline(74.38,14.38)(74.38,0)
\pscustom[]{\psline(71.38,14.38)(77.62,14.38)
\psline(77.62,14.38)(74.5,19.38)
\psbezier(74.5,19.38)(74.5,19.38)(74.5,19.38)
\psline(74.5,19.38)(71.38,14.38)
\closepath}
\rput(74.38,16.38){$\delete$}
\rput[r](73.12,1.25){$\One$}
\end{pspicture}

\end{center}

\subsubsection{Putting it all together}
In a monoidal computer, all of the above operations can be used to move the data around as needed. E.g., if the morphism in Fig.~\ref{fig-String-diagram-equations} represents a computation requiring two inputs $x_1$ and $x_2$ of type $A$, and producing two outputs $z_1$ and $z_2$ of type $C$ and one output $v$ of type $D$, then we can use the data service structure to feed the same value $x$ for both $x_1$ and $x_2$, to filter the values $z_1$ and $z_2$, and to delete $v$. 
\begin{figure}
\def\JPicScale{.9}
\centering
\newcommand{\aah}{A}
\newcommand{\ch}{C}
\renewcommand{\dh}{D}
\newcommand{\ahh}{\scriptstyle (gCD)(fkh)}
\newcommand{\formula}{\scriptstyle (gCh)(\varsigma D)(kfA)}
\newcommand{\eql}{\scriptstyle =}
\newcommand{\delete}{\cun}
\newcommand{\diagonal}{\cmn}
\newcommand{\codiagonal}{\mnd}
\input{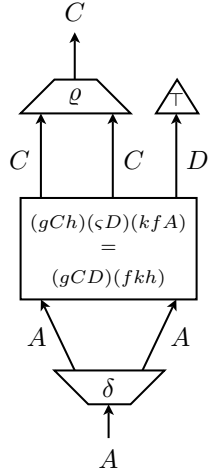}
\caption{Using data service}
\label{fig-data-service}
\end{figure}
This is shown in Fig.~\ref{fig-data-service}. The algebraic expressions for the monoidal morphisms in Fig.~\ref{fig-String-diagram-equations} are here shortened by eliding $\otimes$ within the parentheses and $\circ$ outside, as it is often done when confusion seems unlikely.

\be{defn}
A \emph{data service} in a strict symmetric monoidal category $(\CCC,\otimes, I)$ is a quadruple $(A,\cun,\cmn,\mnd)$ where 
\begin{itemize}
\item $A\in \lvert\CCC\rvert$ is the underlying data type,
\item $\cun: A\to I$ is the \emph{deleting} operation,
\item $\cmn: A \to A\otimes A$ is the \emph{copying} operation
\begin{itemize}
\item such that $A\otimes A \oot \cmn A \tto \cun I$ is a commutative monoid,
\end{itemize}
\item $\mnd: A\otimes A \to A$ is the \emph{filtering\/} operation
\begin{itemize}
\item which makes $A$ into a commutative semigroup.
\end{itemize}
\end{itemize}
The copying and filtering operations are furthermore required to satisfy the \emph{data distribution} conditions
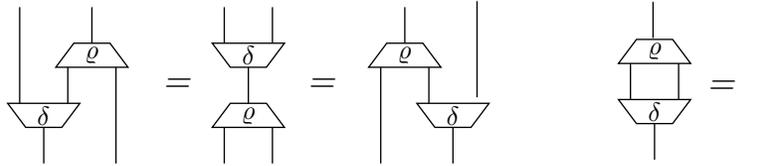
\beq \label{eq-Frobenius}
(A\otimes \mnd)\circ (\cmn\otimes A)\ =\ \cmn \circ \mnd\ =\ (\mnd\otimes A)\circ(A\otimes \cmn) \qquad \qquad\quad \mnd\circ \cmn = \id
\eeq
\begin{center}
\newcommand{\eqls}{$\mbox{\Large =}$}
\newcommand{\comonoid}{\cmn}
\newcommand{\monoid}{\mnd}
\def\JPicScale{.8}
\ifx\JPicScale\undefined\def\JPicScale{1}\fi
\psset{unit=\JPicScale mm}
\psset{linewidth=0.3,dotsep=1,hatchwidth=0.3,hatchsep=1.5,shadowsize=1,dimen=middle}
\psset{dotsize=0.7 2.5,dotscale=1 1,fillcolor=black}
\psset{arrowsize=1 2,arrowlength=1,arrowinset=0.25,tbarsize=0.7 5,bracketlength=0.15,rbracketlength=0.15}
\begin{pspicture}(0,0)(126.25,26.75)
\rput(47,17.75){}
\psline[linewidth=0.22](2.25,9.75)(2.25,25.75)
\psline[linewidth=0.22](14.25,25.75)(14.25,19.75)
\psline[linewidth=0.22](36.25,19.75)(36.25,25.75)
\pspolygon[linewidth=0.22](34.25,19.75)
(46.25,19.75)
(43.25,15.75)
(37.25,15.75)(34.25,19.75)
\psline[linewidth=0.22](10.25,9.75)(10.25,15.75)
\pspolygon[linewidth=0.22](0.25,9.75)
(12.25,9.75)
(9.25,5.75)
(3.25,5.75)(0.25,9.75)
\psline[linewidth=0.22](6.25,5.75)(6.25,-0.25)
\psline[linewidth=0.22](44.25,19.75)(44.25,25.75)
\pspolygon[linewidth=0.22](20.25,15.75)
(8.25,15.75)
(11.25,19.75)
(17.25,19.75)(20.25,15.75)
\rput(14.25,17.75){$\monoid$}
\psline[linewidth=0.22](18.25,-0.25)(18.25,15.75)
\rput(6.25,7.75){$\comonoid$}
\psline[linewidth=0.22](40.25,15.75)(40.25,9.75)
\psline[linewidth=0.22](36.25,-0.25)(36.25,5.75)
\pspolygon[linewidth=0.22](46.25,5.75)
(34.25,5.75)
(37.25,9.75)
(43.25,9.75)(46.25,5.75)
\rput(40.25,7.75){$\monoid$}
\psline[linewidth=0.22](44.25,-0.25)(44.25,5.75)
\rput(40.25,17.75){$\comonoid$}
\rput(28.25,12.75){$\eqls$}
\psline[linewidth=0.22](78.25,10.75)(78.25,26.75)
\psline[linewidth=0.22](66.25,25.75)(66.25,19.75)
\psline[linewidth=0.22](62.25,-0.25)(62.25,15.75)
\pspolygon[linewidth=0.22](68.25,9.75)
(80.25,9.75)
(77.25,5.75)
(71.25,5.75)(68.25,9.75)
\psline[linewidth=0.22](74.25,5.75)(74.25,-0.25)
\pspolygon[linewidth=0.22](72.25,15.75)
(60.25,15.75)
(63.25,19.75)
(69.25,19.75)(72.25,15.75)
\rput(66.25,17.75){$\monoid$}
\psline[linewidth=0.22](70.25,9.75)(70.25,15.75)
\rput(74.25,7.75){$\comonoid$}
\rput(52.25,12.75){$\eqls$}
\rput(114.5,8.38){}
\psline[linewidth=0.22](103.75,10.38)(103.75,16.38)
\pspolygon[linewidth=0.22](101.75,10.38)
(113.75,10.38)
(110.75,6.38)
(104.75,6.38)(101.75,10.38)
\psline[linewidth=0.22](111.75,10.38)(111.75,16.38)
\psline[linewidth=0.22](107.75,6.38)(107.75,0.38)
\pspolygon[linewidth=0.22](113.75,16.38)
(101.75,16.38)
(104.75,20.38)
(110.75,20.38)(113.75,16.38)
\rput(107.75,18.38){$\monoid$}
\rput(107.75,8.38){$\comonoid$}
\psline[linewidth=0.22](126.25,0)(126.25,25.62)
\rput(118.75,12.5){$\eqls$}
\psline[linewidth=0.22](107.5,26.62)(107.5,20.62)
\end{pspicture}

\end{center}
\ee{defn}

\subsection{Main examples and explanations}
 \paragraph{Cartesian services?} The simplest examples of the copying and the deleting operations are given by the cartesian structure, say in the category $\Set$ of sets and functions, where $\otimes$ is the cartesian product $\times$, and $\diag:A\to A\times A$ doubles each $a\in A$ into the pair $<a,a>\in A\times A$, whereas the unit $I$ is the terminal object $\doublone$, and $\bang : A\to \doublone$ deletes all $a\in A$ by mapping them into the unique element $\emptyset \in \doublone$. Indeed, it is easy to see that the monoidal structure of any given category is cartesian precisely when there are \emph{natural\/} copying and deleting operations $(\cmn,\cun)$ on each of its objects. The naturality of the copying and the deleting operations just means that all morphisms preserve them, in the sense that $\cmn_B \circ f = (f\otimes f)\circ \cmn_A$ and $\cun_B\circ f = \cun_A$ hold for all  $A\tto f B$. In terms of Sec.~\ref{sec-homomorphisms}, this means that all morphisms $f\in \CCC(A,B)$ are comonoid homomorphisms. Intuitively, these two preservation properties can be understood as telling that $f$ is single-valued, and that it is total. This is exactly what they mean in the category $\Rel$ of sets and relations.

Adding a data filtering operation $\mnd$ destroys the cartesian structure, because the requirement that $A\otimes A\tto \mnd A$ is a comonoid homomorphism, together with the Frobenius condition, implies that $A$ must be trivial. For this reason, the cartesian structure does not provide a full data service.

\paragraph{Partial functions.} A simple example of full data services can be found in the category $\Pfn$ of sets and \emph{partial\/} functions, with the monoidal structure induced by the cartesian products of sets. This means that the copying and the deleting operations are as described above, but they do not form a cartesian structure with respect to \emph{partial\/} functions, because the requirement $\cun_B\circ f = \cun_A$ just says that the function $f$ must be total. The filtering operation $\mnd:A\otimes A \to A$ maps $<a,a>$ into $a$, and remains undefined on $<a,b>$ when $a\neq b$. 

\paragraph{What is the meaning of the data distribution conditions?} The first equation in \eqref{eq-Frobenius} is known as the \emph{Frobenius condition} \cite{Carboni-Walters,PavlovicD:QMWS}. It allows bringing any well typed expression formed of $\cmn$s and $\mnd$s to a normal form, where all $\cmn$s come after all $\mnd$s. Since the associativity laws make equal, on one hand, all different expressions $\Delta: A\to A^{\otimes n}$, for any fixed $n$, formed of $\cmn$s alone, and on the other hand all different expressions $\nabla: A^{\otimes m}\to A$, for any fixed $m$, formed of $\mnd$s alone,
the effect of these Frobenius normalizations is that any operation composed from $\cmn$s and $\mnd$s boils down to a "spider" in the form $A^{\otimes m}\tto \nabla A\tto \Delta A^{\otimes n}$, with $m$ legs coming in and $n$ legs coming out  \cite{PavlovicD:CQStruct,CoeckeB:picturalism}. The body of such a "spider" is a \emph{data distribution point}. If in a data service in the category $\Pfn$ of partial functions the same value enters a distribution point through all of the strings coming in, then this value will be distributed through all of the strings going out; otherwise, if some of the incoming values are not equal, then no value will come out. 

The second equation in \eqref{eq-Frobenius} implies that the data that can be copied and deleted are \emph{normal}, in the sense that will be spelled out in Prop.~\ref{prop-basis} below. Its logical meaning was analyzed in more detail in \cite[Sec.~4.2]{PavlovicD:Qabs}. Some other logical consequences of the Frobenius conditions, that will be used in the sequel of this work, were analyzed in \cite[Sec.~4.1]{PavlovicD:Qabs}.

%
%

\paragraph{Relations.} 
The category $\Rel$ of sets and binary relations, still with the monoidal structure induced by the cartesian products of sets, clearly contains all data services contained in its subcategory $\Pfn$. The difference is that the filtering operation $\mnd\in \Rel(A\times A, A)$ of any relational data service has a unit $\unt\in \Rel(\doublone, A)$, and thus forms a monoid. This unit does not exist in $\Pfn$. For the data service induced on the set $A$ by the cartesian comonoid $A\times A \oot \diag A \tto \bang \doublone$, this unit is provided by the "chaotic" relation $\unt = \bang^{op} =  \left\{<\emptyset, a> \ |\ a\in A\right\}$, which is obviously not a partial function. 

In general, it can be shown that a data service where the filtering operations form monoids must be self-dual. Indeed, a monoid and a comonoid connected by the Frobenius law form a \emph{Frobenius algebra}, which is a self-dual structure \cite{StreetR:Frobenius}. Frobenius algebras have a rich mathematical theory, and their computational interpretations have been recently recognized in categorical quantum mechanics \cite{PavlovicD:QMWS,PavlovicD:CQStruct,PavlovicD:Qabs}. The Frobenius structure makes its carrier self-dual \cite[Thm.~4.3]{PavlovicD:Qabs}, and a monoidal category where all objects are self-dual is compact \cite{Kelly-Laplaza}.

\paragraph{Vector spaces.} The set based examples of data services share an important property which is not always satisfied: namely that any element $a: I\to A$ can be copied by the copying operation, in the sense that $\cmn\circ a = a\otimes a$. This is not true in general. To understand this, consider the category $\Vect$ of vector spaces and linear operators. A comonoid over a space $A$ can be defined by selecting an orthnormal basis $\Base{\Vect}(A)$ for this space and by defining the linear operator $\cmn\in \Vect(A, A\otimes A)$ by the matrix that sends the basis vectors $|b>\in \Base\Vect(A)$ to the basis vectors $|bb>\in \Base\Vect(A\otimes A)$. Remarkably, it turns out that \emph{every}\/ commutative Frobenius algebra over a vector space $A$ comes about in this way, i.e. that the vectors $\alpha\in A$ satisfying $\cmn\otimes \alpha = \alpha\otimes \alpha$ form an orthogonal family which spans $A$, provided that $A$ is finitely dimensional \cite{PavlovicD:MSCS08}. Furthermore, dropping the unit $\unt$ and relaxing the Frobenius monoid to a semigroup corresponds to dropping the finiteness requirement on the basis \cite{Abramsky-Heunen:H-algs}. A basis of a vector space $A$ can thus be specified entirely in terms of linear operators --- just by specifying a data service on it. 

\subsection{Representing, copying and deleting data}
In a monoidal category $\CCC$, the data of type $A$ are presented as the morphisms $I\to A$, i.e. the elements of $\CCC(A)$. In $\Pfn$, the data values $a\in \Pfn(A)$ are thus the partial functions $\doublone \tto a A$, which are either the elements of the set $A$ in the usual sense, or the empty function $\doublone \tto\emptyset A$. In $\Rel$, the data values $a\in \Rel(A)$ are just the subsets of the set $A$. In $\Vect$ the data values $a\in \Vect(A)$ are the vectors in the vector space $A$.

\subsubsection{Basic data}
A data value $a\in \CCC(A)$ can be copied by a data service 
$(A,\cun,\cmn,\mnd)$ 
if $\cmn \circ a = a\otimes a$. It can be deleted if $\cun \circ a = \id_I$. In terms of Sec.~\ref{sec-homomorphisms}, this means that $a\in \CCC(A)$ can be copied and deleted if and only if the morphism $a:\doublone \to A$ is a homomorphism from the comonoid $I\otimes I \oot = I \tto = I$ to the comonoid  $A\otimes A \oot \cmn  A\tto \cun I$.

\be{defn}\label{def-basic-data}
A \emph{basic} data value, or an \emph{(abstract) element} with respect to a data service $(A, \cun, \cmn,\mnd)$ is a data value $a\in \CCC(A)$ that can be copied and deleted: it corresponds to a comonoid homomorphism $a: I\to A$. 

$\Base\CCC(A,\cun,\cmn,\mnd)$ denotes the set of basic data values with respect to the data service $(A,\cun,\cmn,\mnd)$ in $\CCC$. It is abbreviated to $\Base\CCC(A)$ whenever the confusion is unlikely. This is the \emph{basis} of $\CCC(A)$.

The data values that are not basic are called \emph{mixed}.
\ee{defn}

\subsubsection{Examples} 
We mentioned above that $\Base\Vect(A,\cun,\cmn,\mnd) \subseteq \Vect(A)$ is a basis of the vector space $A$ uniquely determined by the given data service. The vectors in $\Vect(A) \setminus \Base\Vect(A)$ are mixed, and cannot be copied and deleted. 

At the first sight, the situation in the set-based examples seems simpler: the cartesian data services $A\times A \underset{\codiag}{\overset{\diag}{\leftrightarrows}}  A\tto \bang I$ allow copying all data $a\in \Pfn(A)$, and they allow deleting all data except $\emptyset \in \Pfn(A)$. Thus $\Base\Pfn(A) = A$, and the abstract elements of a set $A$ defined in Def.~\ref{def-basic-data} coincide with the usual elements of $A$. In $\Rel$, a subset $a\in \Rel(A)$ can be copied by the cartesian data service if and only if it has at most one element, and it can be deleted if and only if it has at least one element. So $\Base\Rel(A,\bang,\diag,\codiag) = A$ again. However, besides these cartesian data services, the category $\Rel$ admits many \emph{nonstandard\/} data services. This is a consequence of the fact that there are nonstandard commutative Frobenius algebras in $\Rel$, and we have explained above that such algebras are just data services where the filtering semigroup is a monoid. These nonstandard Frobenius algebras in $\Rel$ were analyzed in \cite{PavlovicD:QI09}, where it has been shown that each commutative Frobenius algebra in $\Rel$ corresponds to a partition of $A$ into a disjoint union $A = \coprod_{j\in J} A_j$, where each $A_j$ carries the structure of an abelian group. The standard Frobenius algebras, induced by the cartesian structure, correspond to the special case where all parts $A_j$ are one element sets, with the trivial group structure. This analysis lifts from Frobenius algebras to data services, as it not depend on the units, as explained in \cite{Abramsky-Heunen:H-algs}. A data service $A\otimes A \underset{\mnd}{\overset{\cmn}{\leftrightarrows}}  A\tto \cun I$ in $\Rel$ thus corresponds to a partition of $A$ into abelian groups. It follows that every data service in $\Rel$ is a Frobenius algebra. The group structures over the disjoint parts $A_j\subseteq A$ can be conjoined into a monoid of relations over  $A = \coprod_{j\in J} A_j$ in the form
\[\prooftree
A_j\times A_j \tto{+_j} A_j\oot{o_j} \doublone
\using 
(j\in J)
\justifies
A\ \times A\  \tto{+\ } A\  \oot{o\ } \doublone
\endprooftree
\]
where $+$ is the partial function such that $a+b$ is defined to be $a+_j b$ if and only if $a,b\in A_j$ for some $j\in J$, otherwise undefined; and where $o$ is the multivalued relation, relating the unique element of $\doublone$ with $o_j \in A_j$ for all $j\in J$. In the corresponding data service  $(A,\cun,\cmn,\mnd)$, the copying and deleting comonoid $A\times A \oot \cmn A \tto \cun \doublone$ consists of the opposite relations of the filtering monoid $A\times A \tto \mnd A \oot \unt \doublone$, which is just this abelian group structure, with $\mnd = +$ and $\unt = o$. It is easy to see that a abstract element of $A$ must support this group structure, and that they are thus just the underlying sets of the partition, i.e.
\bear
\Base{\Rel}(A) & = & \left\{A_j\subseteq A\ |\  j\in J\right\}
\eear
All other subsets  $a\subseteq A$ correspond to mixed data, which cannot be copied and deleted. This lifts an abstract version of the basis decomposition from vector spaces and linear operators to sets and relations, which turn out to support toy models of quantum computation \cite{PavlovicD:Qabs}.

In fact, a data service structure can even be viewed as a rudimentary Hilbert space structure, which will be useful in modeling randomized computation.

\subsubsection{Functions and mixtures}
\be{defn}\label{def-functions}
The \emph{pure}\/ morphisms, or \emph{(basic) functions} with respect to data services $(A, \cun_A, \cmn_A,\mnd_A)$ and $(B, \cun_B, \cmn_B,\mnd_B)$ are the morphisms $f\in \CCC(A,B)$ that correspond to comonoid homomorphisms $f: A\to B$. 

The set of pure morphisms with respect the given data services on $A$ and $B$ are written $\Base\CCC(A,B)$. This is the \emph{basis} of $\CCC(A,B)$.

The morphisms that are not pure are called \emph{mixed}.
\ee{defn}

\paragraph{Remark.} The functions are obviously closed under composition, and they include identities, so they form a category. The category $\CCC_\times$ of comonoids in $\CCC$ and functions (comonoid homomorphisms) between them is the cofree cartesian category over $\CCC$, i.e. it comes with a  functor $\CCC_\times \to \CCC$, forgetting the comonoid structure \cite{Fox}.

\paragraph{Explanation.} A partial function in $\Pfn$ is pure if and only if it is total. A possible computational interpretation that will become clear in the sequel is that a function that is not total cannot be deleted by a data service, because it cannot decide whether this function will halt or not. A relation in $\Rel$ is pure with respect to the standard cartesian data services if and only if it is a function in the usual sense, i.e., it is a  total and single-valued relation. The totality means that it preserves the cartesian comonoid unit; the single-valuedness means that it preserves the comonoid diagonal.  A possible computational interpretation in terms of relations as the denotations of nondeterministic computations is that a data service cannot run in parallel several copies of a relation that is not single-valued, because the different copies of the same relation may make different nondeterministic choices at runtime. A linear operator $f\in \Vect(A,B)$ is pure with respect to the data services with the bases $\Base\Vect(A)$ and $\Base\Vect(B)$ if and only if it is induced by a function $\Base f :\Base \Vect(A) \to \Base\Vect(B)$. Clearly, most linear operators are mixtures. The relational mixtures with respect to nonstandard data services will play an important role in the sequel.

\begin{prop}
Let $A$ and $B$ be data services in $\Rel$ induced by the abelian group decompostions $A = \coprod_{j\in J} A_j$ and $B = \coprod_{k\in K} B_k$. A relation $f\in \Rel(A,B)$ if and only if there is a function $\varphi:J\to K$ and $f$ decomposes into a disjoint union $f = \coprod_{j\in J} f_j$ where each $f_j \in \Rel\left(A_j, B_{\varphi(j)}\right)$ is a chaotic relation with $af_j b$ for all $a\in A_j$ and all $b\in B_{\varphi(j)}$.
\end{prop}

\subsection{Convolution, norm, inner product}
\be{defn}
Any data service on $A$ induces the operations
\begin{itemize}
\item convolution $\star \ :\  \CCC(A,B) \times \CCC(A,B) \longrightarrow  \CCC(A,B)$ with
\bear
f\star g  & = &  \mnd_B\circ (f\otimes g) \circ \cmn_A
\eear
\item norm $\lvert -\rvert \  :\ \CCC(A) \longrightarrow  \CCC(I)$ with
\bear
 \lvert a \rvert & = & \cun_A \circ a
\eear 
\item inner product $<-|-> \ :\  \CCC(A) \times \CCC(A) \longrightarrow  \CCC(I)$ with
 \bear
<a | b> & = &  \cun_A \circ \mnd _A \circ (a\otimes b)
\eear
\end{itemize}

\begin{figure}
\begin{minipage}[b]{0.3\linewidth}
\centering
\newcommand{\one}{f}
\newcommand{\two}{g}
\renewcommand{\Source}{A}
\newcommand{\Target}{B}
\def\JPicScale{1}
\input{PIC/convolution}
\caption{Convolution}
\label{def-convolution}
\end{minipage}
\hspace{.3cm}
\begin{minipage}[b]{0.3\linewidth}
\centering
\newcommand{\one}{a}
\def\JPicScale{1}
\ifx\JPicScale\undefined\def\JPicScale{1}\fi
\psset{unit=\JPicScale mm}
\psset{linewidth=0.3,dotsep=1,hatchwidth=0.3,hatchsep=1.5,shadowsize=1,dimen=middle}
\psset{dotsize=0.7 2.5,dotscale=1 1,fillcolor=black}
\psset{arrowsize=1 2,arrowlength=1,arrowinset=0.25,tbarsize=0.7 5,bracketlength=0.15,rbracketlength=0.15}
\begin{pspicture}(0,0)(6.25,27.5)
\psline{->}(3.12,11.88)(3.12,24.38)
\rput(3.12,10.62){$\one$}
\pscustom[]{\psline(6.25,24.38)(0,24.38)
\psline(0,24.38)(3.12,27.5)
\psbezier(3.12,27.5)(3.12,27.5)(3.12,27.5)
\psline(3.12,27.5)(6.25,24.38)
\closepath}
\pscustom[]{\psline(6.25,11.88)(0,11.88)
\psline(0,11.88)(3.12,8.12)
\psbezier(3.12,8.12)(3.12,8.12)(3.12,8.12)
\psline(3.12,8.12)(6.25,11.88)
\closepath}
\end{pspicture}
\caption{Norm}
\label{def-norm}
\end{minipage}
\hspace{.3cm}
\begin{minipage}[b]{0.3\linewidth}
\centering
\newcommand{\one}{a}
\newcommand{\two}{b}
\def\JPicScale{1.1}
\ifx\JPicScale\undefined\def\JPicScale{1}\fi
\psset{unit=\JPicScale mm}
\psset{linewidth=0.3,dotsep=1,hatchwidth=0.3,hatchsep=1.5,shadowsize=1,dimen=middle}
\psset{dotsize=0.7 2.5,dotscale=1 1,fillcolor=black}
\psset{arrowsize=1 2,arrowlength=1,arrowinset=0.25,tbarsize=0.7 5,bracketlength=0.15,rbracketlength=0.15}
\begin{pspicture}(0,0)(16.25,27.5)
\psline{->}(3.12,10)(3.12,15.62)
\pspolygon(16.25,15.62)
(0,15.62)
(5,20.62)
(11.25,20.62)(16.25,15.62)
\psline{->}(8.12,20.62)(8.12,24.38)
\psline{->}(13.12,10)(13.12,15.62)
\rput(3.12,8.75){$\one$}
\rput(13.12,8.75){$\two$}
\pscustom[]{\psline(11.25,24.38)(5,24.38)
\psline(5,24.38)(8.12,27.5)
\psbezier(8.12,27.5)(8.12,27.5)(8.12,27.5)
\psline(8.12,27.5)(11.25,24.38)
\closepath}
\pscustom[]{\psline(16.25,10)(10,10)
\psline(10,10)(13.12,6.25)
\psbezier(13.12,6.25)(13.12,6.25)(13.12,6.25)
\psline(13.12,6.25)(16.25,10)
\closepath}
\pscustom[]{\psline(6.25,10)(0,10)
\psline(0,10)(3.12,6.25)
\psbezier(3.12,6.25)(3.12,6.25)(3.12,6.25)
\psline(3.12,6.25)(6.25,10)
\closepath}
\end{pspicture}
\caption{Inner product}
\label{def-inner-product}
\end{minipage}
\end{figure}
\ee{defn}

\paragraph{Examples.} In $\Vect$, these operations take their usual meaning: the abstract inner product is the usual inner product, the norm is the $\ell_1$-norm, and the convolution of the matrices $f$ and $g$ is the entry-wise multiplication $(f\star g)_i = f_i \cdot g_i$, so that $<a |b> = \lvert a \star b \rvert$, where the convolution is taken for $A = I$. It is easy to see that this holds in general.

In $\Pfn$, $f\star g = f\cap g$ is the intersection, so that $f\star f = f$ hods for all $f$. This is also true with respect to the standard, i.e. cartesian data services in $\Rel$.

\subsubsection{Abstract relations}
\be{defn}\label{def-relations}
A morphism $f\in \CCC(A,B)$ is an \emph{(abstract) relation} with respect to some given data service on $A$ and $B$ if $f\star f = f$.
\ee{defn}

\begin{prop}
The convolution operation is always associative and commutative, and thus makes $\CCC(A,B)$ into a commutative semigroup relative to any given data services on $A$ and $B$. Abstract relations form a subsemilattice of each of these semigroups. 
\end{prop}

\paragraph{Examples.} All morphisms in $\Pfn$ are abstract relations. All morphisms in $\Rel$ are abstract relations with respect to the standard, cartesian data services. The situation is more interesting with respect to the nonstandard data services. 

\begin{prop}
Let $A$ and $B$ be data services in $\Rel$ induced by the abelian group decompostions $A = \coprod_{j\in J} A_j$ and $B = \coprod_{k\in K} B_k$.  Then $f\in \Rel(A,B)$ is an abstract relation with respect to the induced convolution operation $\star: \Rel(A,B)\times \Rel(A,B) \to \Rel(A,B)$ if and only if there is a partial bijection $\varphi:J\to K$ and $f$ decomposes into a disjoint union $f = \coprod_{j\in J} f_j$, where $f_j=\emptyset$ if $\varphi(j)$ is undefined, and otherwise $f_j \in \Rel\left(A_j, B_{\varphi(j)}\right)$ is a congruence, in the sense that $ufx \wedge vfy \Longrightarrow (u+v)f(x+y)\wedge (-u)f(-x)$.
\end{prop}

\paragraph{More examples.} Since the convolution in $\Vect$ is the entry-wise multiplication of the matrices,  a linear operator $f\in \Vect(A,B)$ is an abstract relation with respect to the bases $\Base\Vect(A)$ and $\Base\Vect(B)$ if and only if the entries of its matrix representation are idempotent, i.e. they must all be 0s or 1s. While such matrices can be naturally viewed as binary relations between the basis elements, their matrix composition in $\Vect$ is not the usual relational composition. This shows that abstract relations are in general not closed under composition. This was discussed in \cite{PavlovicD:CQStruct}. 

\paragraph{Comment.} Although providing some data services over all objects of $\CCC$ makes all of its hom-set $\CCC(A,B)$ into a semigroups, this structure is generally not preserved by the composition, and therefore it does not make $\CCC$ into a semigroup enriched category \cite{KellyM:Enriched}. 

\subsubsection{Scalars}\label{sec-scalar-monoid}
As mentioned in Sec.~\ref{Sec-cats}, the "elements" of the tensor unit $I$, i.e. the elements of the set $\CCC(I) = \CCC(I,I)$ are called \emph{scalars}. In general, the monoidal structure comes with an isomorphism $I\cong I\otimes I$, which provides $I$ with a canonical Frobenius algebra structure, and thus a data service and a convolution operation. In a \emph{strict}\/ monoidal category, these isomorphisms are identities, i.e. $I=I\otimes I$ holds on the nose. For the scalars $\alpha,\beta\in \CCC(I)$, the strictness assumption and the definition of the convolution together imply
\[ \alpha\star \beta\   =\  \alpha \otimes \beta\ =\  \alpha \circ \beta
\] 
Hence the commutative monoid $(\CCC(I), \star, \id_I)$. We usually elide its operation, and reduce all of the above expressions to $\alpha\beta$.

\subsubsection{Bases are orthonormal}
\be{defn}
A data value $a\in \CCC(A)$ is\/ \emph{normal} if $<a|a> = \id_I$. Data values $a,b\in \CCC(A)$ are\/ \emph{orthogonal} if $<a|b>^2 = <a|b>$. A set $S \subseteq \CCC(A)$ is\/ \emph{orthonormal} if all of its elements are normal and any pair is orthogonal.
\ee{defn}

\be{prop}\label{prop-basis}
The basis $\Base\CCC(A)$ of any type $A$ is orthonormal.
\ee{prop}

The proof is left as an easy exercise in diagrammatic reasoning.

\section{Monoidal computer}\label{Sec-Moncom}
The idea of a monoidal computer is that it should provide a data service $\CCC$ where all morphisms $f\in \CCC(A,B)$ are computable. One way to say that $f$ is computable is to require that that there is a program $p$ which encodes the computation $f$. Since a program needs to be manipulated, transformed and composed, it should be a basic data value, that can be copied and deleted. We thus require that for any pair of types $L,M$ in $\CCC$ there is a surjective "program execution" operation $u:\Base\CCC(\Nn) \epi \CCC(L,M)$ that interprets the basic data values $p\in \Base\CCC(\Nn)$ as programs and assigns to them the corresponding computations $u(p)\in \CCC(L,M)$. 

\be{defn}\label{def-moncomp}
A \emph{(basic) monidal computer}\/ is a data service $\CCC$ which has:
\begin{itemize}
\item a \emph{universal data type} $\Nn$: it generates all types as its tensor powers, i.e. for every $M\in \lvert \CCC\rvert$ there is $m\geq 0$ such that $M = \Nn^{\otimes m}$
\item \emph{universal evaluators}\/ $\UK^M_L\in \CCC(\Nn\otimes L, M)$, indexed by $L,M\in |\CCC|$: for every computation $f\in \CCC(L,M)$ there is an element $p\in \Base\CCC(\Nn)$, called a \emph{program\/} for $f$, such that
\bea\label{eq-universal-evaluator}
 \def\JPicScale{1}\newcommand{\aah}{L}\renewcommand{\dh}{M}\newcommand{\ahh}{f} 
\ifx\JPicScale\undefined\def\JPicScale{1}\fi
\psset{unit=\JPicScale mm}
\psset{linewidth=0.3,dotsep=1,hatchwidth=0.3,hatchsep=1.5,shadowsize=1,dimen=middle}
\psset{dotsize=0.7 2.5,dotscale=1 1,fillcolor=black}
\psset{arrowsize=1 2,arrowlength=1,arrowinset=0.25,tbarsize=0.7 5,bracketlength=0.15,rbracketlength=0.15}
\begin{pspicture}(0,0)(8,13.75)
\rput(5,13.75){$\dh$}
\psline{->}(5,-8.25)(5,-1.25)
\pspolygon[](2,4.75)(8,4.75)(8,-1.25)(2,-1.25)
\psline{->}(5,4.75)(5,11.75)
\rput(5,1.75){$\ahh$}
\rput(5,-11.25){$\aah$}
\end{pspicture}
 & = & \def\JPicScale{.9}\newcommand{\aah}{p}\renewcommand{\dh}{M}\newcommand{\ahh}{\UK^M_L}\newcommand{\bhh}{L} 
\ifx\JPicScale\undefined\def\JPicScale{1}\fi
\psset{unit=\JPicScale mm}
\psset{linewidth=0.3,dotsep=1,hatchwidth=0.3,hatchsep=1.5,shadowsize=1,dimen=middle}
\psset{dotsize=0.7 2.5,dotscale=1 1,fillcolor=black}
\psset{arrowsize=1 2,arrowlength=1,arrowinset=0.25,tbarsize=0.7 5,bracketlength=0.15,rbracketlength=0.15}
\begin{pspicture}(0,0)(20,13.75)
\rput(17,13.75){$\dh$}
\psline{->}(5,-6.25)(5,0.75)
\pspolygon[](2,4.75)(20,4.75)(20,-1.25)(2,-1.25)
\psline{->}(17,4.75)(17,11.75)
\rput(16,1.75){$\ahh$}
\rput(5,-8.12){$\aah$}
\psline{->}(17,-8.25)(17,-1.25)
\rput(17,-11.25){$\bhh$}
\pspolygon(2.5,-6.25)
(7.5,-6.25)
(5.01,-11.88)(2.5,-6.25)
\pspolygon(3.75,3.12)
(6.25,3.12)
(5,0.62)(3.75,3.12)
\end{pspicture}

\eea

\vspace{3\baselineskip}

\item \emph{partial evaluators}\/ $\SK_{MN}\in \Base\CCC( \Nn\otimes M, \Nn)$, indexed by $M,N\in \lvert \CCC\rvert$, such that

\bea\label{eq-partial-evaluator}
\def\JPicScale{.9}\renewcommand{\dh}{\Nn}\newcommand{\ahh}{\UK_N^L}\newcommand{\shh}{\SK_{MN}}\newcommand{\bhh}{M}\newcommand{\chh}{N}\newcommand{\Lhh}{L} 
\ifx\JPicScale\undefined\def\JPicScale{1}\fi
\psset{unit=\JPicScale mm}
\psset{linewidth=0.3,dotsep=1,hatchwidth=0.3,hatchsep=1.5,shadowsize=1,dimen=middle}
\psset{dotsize=0.7 2.5,dotscale=1 1,fillcolor=black}
\psset{arrowsize=1 2,arrowlength=1,arrowinset=0.25,tbarsize=0.7 5,bracketlength=0.15,rbracketlength=0.15}
\begin{pspicture}(0,0)(30,27.5)
\rput(3,9){$\dh$}
\psline{->}(5,-8)(5,0.62)
\pspolygon[](2,4.75)(20,4.75)(20,-1.25)(2,-1.25)
\psline{->}(5,3.12)(5,14.38)
\psline{->}(16.88,-8)(16.88,-1.25)
\rput(16.88,-11.25){$\bhh$}
\pspolygon[](1.88,18.62)(30,18.62)(30,12.62)(1.88,12.62)
\rput(25,15.62){$\ahh$}
\rput(15.62,1.88){$\shh$}
\psline{->}(26.25,-8.12)(26.25,12.5)
\rput(26.25,-11.25){$\chh$}
\rput(5,-11){$\dh$}
\psline{->}(26.25,18.75)(26.25,25)
\rput(26.25,27.5){$\Lhh$}
\pspolygon(3.75,3.12)
(6.25,3.12)
(5,0.62)(3.75,3.12)
\pspolygon(3.75,16.88)
(6.25,16.88)
(5,14.38)(3.75,16.88)
\end{pspicture}
 & = & \def\JPicScale{.9}\renewcommand{\dh}{\Nn}\newcommand{\ahh}{\UK_{M\otimes N}^L}\newcommand{\shh}{\SK}\newcommand{\bhh}{M}\newcommand{\chh}{N}\newcommand{\Lhh}{L} 
\ifx\JPicScale\undefined\def\JPicScale{1}\fi
\psset{unit=\JPicScale mm}
\psset{linewidth=0.3,dotsep=1,hatchwidth=0.3,hatchsep=1.5,shadowsize=1,dimen=middle}
\psset{dotsize=0.7 2.5,dotscale=1 1,fillcolor=black}
\psset{arrowsize=1 2,arrowlength=1,arrowinset=0.25,tbarsize=0.7 5,bracketlength=0.15,rbracketlength=0.15}
\begin{pspicture}(0,0)(30,27.5)
\rput(5,-11.25){$\dh$}
\psline{->}(5,-8.12)(5,8.12)
\psline{->}(21.25,-8)(21.25,6.25)
\rput(21.25,-11.25){$\bhh$}
\pspolygon[](1.88,12.38)(30,12.38)(30,6.38)(1.88,6.38)
\rput(23.75,9.38){$\ahh$}
\psline{->}(26.25,-8.12)(26.25,6.25)
\rput(26.25,-11.25){$\chh$}
\psline{->}(23.75,12.5)(23.75,25)
\rput(23.75,27.5){$\Lhh$}
\pspolygon(3.75,10.62)
(6.25,10.62)
(5,8.12)(3.75,10.62)
\end{pspicture}

\eea

\vspace{3\baselineskip}
\end{itemize}
\ee{defn}

%
\paragraph{Notation and intuition for the universal evaluators.} The universal evaluators provide a monoidal view of the idea of a programming language. In theory of computability (e.g. \cite{RogersH:book}), this idea is formalized by the enumerations of computable functions. The programs thus boil down to numeric indices, and their executions are denoted using the \emph{Kleene's brackets}\/ $\{-\}:\NNn \epi \Pfn(\NNn,\NNn)$, so that $\{p\}:\NNn\pfn \NNn$ represents the computation induced by the program $p$. It is thus natural to use the well-known notiation $\deco p$  as the abbreviation for $\UK_L^M\circ (p\otimes L)$. The other way around, it is also convenient to have a notation for a program corresponding to a computation; so we generically write $\enco f$ for an arbitrary program that encodes a given computation $f$. In the monoidal computer formalism, these conventions thus mean
\vspace{-.5\baselineskip}
\[
\def\JPicScale{1.2}
\newcommand{\aah}{}\renewcommand{\dh}{}\newcommand{\ahh}{\deco p} 
\ifx\JPicScale\undefined\def\JPicScale{1}\fi
\psset{unit=\JPicScale mm}
\psset{linewidth=0.3,dotsep=1,hatchwidth=0.3,hatchsep=1.5,shadowsize=1,dimen=middle}
\psset{dotsize=0.7 2.5,dotscale=1 1,fillcolor=black}
\psset{arrowsize=1 2,arrowlength=1,arrowinset=0.25,tbarsize=0.7 5,bracketlength=0.15,rbracketlength=0.15}
\begin{pspicture}(0,0)(8,13.75)
\rput(5,13.75){$\dh$}
\psline{->}(5,-8.25)(5,-1.25)
\pspolygon[](2,4.75)(8,4.75)(8,-1.25)(2,-1.25)
\psline{->}(5,4.75)(5,11.75)
\rput(5,1.75){$\ahh$}
\rput(5,-11.25){$\aah$}
\end{pspicture}

\ = \ 
\renewcommand{\aah}{p}\renewcommand{\dh}{}\renewcommand{\ahh}{}\newcommand{\bhh}{}  
\qquad \qquad \qquad\qquad 
\renewcommand{\aah}{} \renewcommand{\ahh}{f} 
\ifx\JPicScale\undefined\def\JPicScale{1}\fi
\psset{unit=\JPicScale mm}
\psset{linewidth=0.3,dotsep=1,hatchwidth=0.3,hatchsep=1.5,shadowsize=1,dimen=middle}
\psset{dotsize=0.7 2.5,dotscale=1 1,fillcolor=black}
\psset{arrowsize=1 2,arrowlength=1,arrowinset=0.25,tbarsize=0.7 5,bracketlength=0.15,rbracketlength=0.15}
\begin{pspicture}(0,0)(8,13.75)
\rput(5,13.75){$\dh$}
\psline{->}(5,-8.25)(5,-1.25)
\pspolygon[](2,4.75)(8,4.75)(8,-1.25)(2,-1.25)
\psline{->}(5,4.75)(5,11.75)
\rput(5,1.75){$\ahh$}
\rput(5,-11.25){$\aah$}
\end{pspicture}
\ = \ 
\renewcommand{\aah}{\scriptstyle \ulcorner \! f\! \urcorner} \renewcommand{\ahh}{}  
\]

\vspace{3\baselineskip}
With these notations, Eq.~\eqref{eq-universal-evaluator} becomes 
\bear
\deco{\enco{f}} & = &  f
\eear
The dual equation $\enco{\deco p} = p$ is usually not satisfied. It characterizes the special family of computers which happen to be \emph{extensional\/} in the strong sense that each computation corresponds to a \emph{unique\/} program. An example of such a computer will be mentioned below.

\paragraph{Convention.} To simplify geometric reasoning, in diagrams we often omit any redundant labels and brackets, and even denote the program and the corresponding computation by the same name, whenever the distinction between the two is graphically obvious.

\begin{prop}\label{prop-characterization}
Let $\CCC$ be a symmetric monoidal category with data services, and such that every object $M\in |\CCC|$ is in the form $M = \Nn^{\otimes m}$ for some $m\geq 0$. A specification of the universal evaluators and of the partial evaluators as in Def.~\ref{def-moncomp} is equivalent to a specification, for every $L,M\in \lvert \CCC\rvert$, of a family of surjections
\bea\label{nat-surjections}
\gamma^{LM}_X\ :\ \Base\CCC(X,\Nn) &\epi & \CCC(X\otimes L, M)
\eea
natural in $X$. The correspondence of the families $\gamma^{LM}$  and the universal evaluators $\UK^M_L$ is one-to-one, but each such couple there may be different choices of the corresponding partial evaluators $\SK_{LN}$.
\end{prop} 

\bpr
Given a natural family of surjections \eqref{nat-surjections}, define
\begin{itemize}
\item universal evaluators $\UK_L^M = \gamma_\Nn^{LM}(\id_\Nn)$
\item partial evaluators $\SK_{MN}$ such that $\gamma^{NL}_{\Nn \otimes M}(\SK_{MN}) = \UK^L_{M\otimes N}$
\end{itemize}
The partial evaluators are not unique, but they exist because each $\gamma^{NL}_X$ is surjective, so $\UK^L_{M\otimes N}$ must be in the image of $\gamma^{NL}_{\Nn \otimes M}$. The naturality of $\gamma^{LM}_X$ in $X$ implies that the following squares commute
\medskip

\begin{center}
\newcommand{\one}{\scriptstyle f}
\newcommand{\two}{\scriptstyle g}
\newcommand{\three}{\scriptstyle \gamma^{LM}_\Nn}
\newcommand{\four}{\scriptstyle \gamma^{LM}_I}
\newcommand{\five}{\scriptstyle -\circ p}
\newcommand{\six}{\scriptstyle -\circ (p\otimes L)}
\newcommand{\seven}{\scriptstyle \gamma^{NL}_\Nn}
\newcommand{\eight}{\scriptstyle -\circ \SK_{MN}}
\newcommand{\nine}{\scriptstyle -\circ (\SK_{MN}\otimes M)}
\newcommand{\ten}{\scriptstyle \gamma^{NL}_{\Nn\otimes M}}
\newcommand{\UpOne}{\scriptstyle \Base\CCC(\Nn,\Nn)}
\newcommand{\UpTwo}{\scriptstyle \scriptstyle \CCC(\Nn\otimes L, M)}
\newcommand{\UpThree}{\scriptstyle \scriptstyle \CCC(\Nn\otimes N, L)}
\newcommand{\DownOne}{\scriptstyle \Base\CCC(I,\Nn)}
\newcommand{\DownTwo}{\scriptstyle \CCC(L,M)}
\newcommand{\DownThree}{\scriptstyle \Base\CCC(\Nn\otimes M,\Nn)}
\newcommand{\DownFour}{\scriptstyle \CCC(\Nn\otimes  M\otimes N,L)}
\def\JPicScale{0.7}
\ifx\JPicScale\undefined\def\JPicScale{1}\fi
\psset{unit=\JPicScale mm}
\psset{linewidth=0.3,dotsep=1,hatchwidth=0.3,hatchsep=1.5,shadowsize=1,dimen=middle}
\psset{dotsize=0.7 2.5,dotscale=1 1,fillcolor=black}
\psset{arrowsize=1 2,arrowlength=1,arrowinset=0.25,tbarsize=0.7 5,bracketlength=0.15,rbracketlength=0.15}
\begin{pspicture}(0,0)(130.62,35.62)
\rput(5,32.5){$\UpOne$}
\rput(45.62,32.5){$\UpTwo$}
\rput(5,2.5){$\DownOne$}
\psline{<-}(35,32.5)(12.5,32.5)
\psline{<-}(36.88,2.5)(12.5,2.5)
\psline{->}(5,29.38)(5,6.88)
\psline{->}(45.62,29.38)(45.62,6.88)
\rput(45.62,2.5){$\DownTwo$}
\rput(23.75,35.62){$\three$}
\rput(23.75,-0.62){$\four$}
\rput[r](3.75,18.12){$\five$}
\rput[l](46.89,18.12){$\six$}
\rput(88.75,31.88){$\UpOne$}
\psline{<-}(118.75,31.88)(96.25,31.88)
\psline{<-}(116.25,1.88)(98.75,1.88)
\psline{->}(88.75,28.76)(88.75,6.25)
\psline{->}(129.38,28.76)(129.38,6.25)
\rput(88.75,1.88){$\DownThree$}
\rput(130.62,1.88){$\DownFour$}
\rput[r](87.5,18.12){$\eight$}
\rput[l](130,18.12){$\nine$}
\rput(107.5,-1.25){$\ten$}
\rput(129.38,32.5){$\UpThree$}
\rput(106.88,35.62){$\seven$}
\end{pspicture}

\end{center}

The commutativity of the right-hand square gives condition \eqref{eq-partial-evaluator}. The commutativity of the left-hand square means that $\gamma^{LM}_I (p) = \UK^M_L\circ (p\otimes L)$. Condition \eqref{eq-universal-evaluator} thus follows from the assumption that $\gamma^{MN}_I$ is a surjection.

The other way around, given a family of universal evaluators $\UK_L^M$, define 
\bear
\gamma^{LM}_{X}(q) & = & \UK^M_L\circ(q\otimes L)
\eear 
This is easily seen to give a family natural for the functions in and out of $X$. To prove that $\gamma^{LM}_{X}$ is surjective for every $X\in |\CCC|$, we proceed by induction in $i$ where $X= \Nn^{\otimes i}$. Condition \eqref{eq-universal-evaluator} says that the component $\gamma^{LM}_I$ is surjective, and thus gives the base case $i=0$. Condition \eqref{eq-partial-evaluator} gives the inductive step.
\epr

\subsection{Examples}
\paragraph{Computable partial functions and relations.} In the standard model of monoidal computer, the data are represented as bitstrings, and the computations are the computable partial functions, as implemented, say, by deterministic Turing machines. More precisely, let $\Nn = \{0,1\}^\ast$ be the set of all finite binary strings. Let $\CCC$ be the category consisting of 
\begin{itemize}
\item \emph{objects:}\/ cartesian powers of $\{0,1\}^\ast$
\item \emph{morphisms:}\/ computable partial functions.
\end{itemize}
This is a subcategory of the category $\Pfn$ of all sets and all partial functions, so the monoidal structure and the data services are the same, since the copying, deleting and comparing operations are computable. Using \emph{nondeterministic}\/ Turing machines would lead to computable \emph{relations}, and to the monoid computer contained in the category of sets and relations.

The three parts of the definition of monoid computer correspond to three conceptual pillars of theory of computation:
\begin{itemize}
\item The requirement that all types are the tensor powers of the universal data type says that all data are presented as tuples of bitstrings. 
\item The universal evaluation operations $\UK_M^N$ correspond to the general purpose computers, or more formally to the universal Turing machines with $m$ input tapes and $n$ output tapes, where $M = \Nn^{\otimes m}$ and $N = \Nn^{\otimes n}$. 
\item The partial evaluation operations $\SK_{MN}$ correspond to Kleene's smn-functions, or to the partial evaluators used in programming. 
\end{itemize}
These ideas are discussed in detail in many computability theory textbooks and monographs, e.g. \cite{RogersH:book}. The natural surjection spelled out in Proposition~\ref{prop-characterization} is a categorical view of the intensional enumerations of computable functions, which pervade the computability theory books and proofs. Scott's domain theory tightens, in a sense, these intensional enumerations into an extensional isomorphism between a domain of programs and of a domain of the functions encoded by these programs, which leads us to the next example.

\paragraph{Extensional monoidal computer.} Consider a category of domains for denotational semantics, e.g. the category of continuous lattices $\sf CLat$, or of continuous partial orders $\sf Cpo$ \cite{MisloveM:compendium,Abramsky-Jung:domains}. These categories are cartesian closed, and the cartesian structure gives a canonical comonoid structure $X\times X\oot \cmn X \tto \cun \doublone$ on every $X$. For pointed domains, i.e. those with the least element, this structure extends to a full data service, with idempotent convolution. To form a monoid computer, we need a pointed \emph{reflexive\/} domain $\Nn$. The reflexivity here implies that there is an isomorphism $\Nn \underset{v}{\overset{u}\rightleftarrows} \Nn^\Nn$. In other words, for every $f:\Nn\to \Nn$ there is $p = v(f)\in \Nn$ with $f = u(p)$.  

The \emph{extensional\/} monoid computer is defined to be the full subcategory $\CCC$ spanned by the finite powers $\Nn^m$, $m\geq 0$, of a nontrivial reflexive object $\Nn$ in a category of domains with the bottom. The nontriviality assumption means that $\Nn$ contains the discrete set of natural numbers $\omega = \{0,1,2,\ldots \} \subseteq \Nn$. Now the assumptions that $\Nn^\Nn \cong \Nn$ and that $n \subseteq \Nn$ for every $n=\{0,1,\ldots, n-1\}$ together imply that for every $M = \Nn^m$ and $N = \Nn^n$, $m,n\geq 0$ there is a surjection
\bea\label{eq-ext-u}
\Nn & \eepi{\UK_M^N} & N^M
\eea
derived from
\begin{itemize}
\item  $\Nn^{M} = \Nn^{\left(\Nn^m\right)} \cong \Nn$, which is itself derived from $\Nn^{\Nn\times \Nn} \cong \left(\Nn^\Nn\right)^\Nn \cong \Nn^\Nn \cong \Nn$;
\item $N^M = \left(\Nn^n\right)^M \cong \left(\Nn^M\right)^n \cong \Nn^n \epi \Nn^\Nn\cong \Nn$, which lifts $n\hookrightarrow \Nn$.
\end{itemize}
Checking that $\UK_M^N$ in \eqref{eq-ext-u} gives a universal evaluator is straightforward. The partial evaluators are constructed as in Prop.\ref{prop-characterization}. The extensionality means that every computation $f\in N^M$ is represented by a unique progam $p\in \Nn$. This is clear if we only look at the computations with a single input, as it is expressed by the fact that $\UK: \Nn \to \Nn^\Nn$ is an isomorphism. Capturing the computations with finitely many inputs and outputs requires data services, which complicate the picture, and make the universal evaluators $\UK_M^N: \Nn \to N^M$ into mere surjections. A closer inspection of the above construction of $\UK_M^N$ shows that it assigns a unique program to each computation of $m$ inputs and $n$ outputs, and maps the programs of other arities to the bottom.

\paragraph{Quantum computer} can be viewed as a monoidal computer in the category $\Vect_\CCc$ of complex vector spaces and linear operators, with the data services as described in Sec.~\ref{Sec-data-services}. The universal data type is the 2-dimensional vector space $\CCc^2$, which plays the role of cogenerator, similar to the role of the 2-element set in the category $\Rel$. The universal and the partial evaluators are the evaluators of the Deutsch-Turing machines \cite{DeutschD:1985,Bernstein-Vazirani:1993}. The evaluators and their programs have classical descriptions, which in the framework of the monoidal computer formally means that they are basic data, as required by Definitions~\ref{def-basic-data} and \ref{def-moncomp}. On the other hand, the main feature of the quantum computer is that it processes mixed data, and that such processing can be used to execute many computational threads in parallel, with a low computational overhead. The gains from this feature will be captured in monoidal computers in terms of the space complexity.  

A physically more realistic view of quantum computation requires factoring out the irrelevant phases, and quotienting $\CCc^2$ to the data type of qubits. This leads to significantly more complicated notions of quantum operation, and of quantum computation  \cite{Nielsen-Chuang:book}. However, the relevant features of these refined notions do not seem to be limited to their standard vector based realizations, as recognized already by von Neumann \cite{RedeiM:why}. A categorical analysis of quantum computation allows distinguishing the essential structural components, while hiding the inessential implementation details of vector spaces \cite{Coecke-Duncan:long}, and even helps constructing the so-called \emph{toy models}, used to study the dependencies between those components \cite{SpekkensR:toy,Barnum-Wilce:teleportation,Coecke-Edwards,PavlovicD:QPL09}. There are thus many nonstandard quantum computers, some of them reducing the qubit computations all the way to boolean relations \cite{PavlovicD:QI09,PavlovicD:Qabs}.

\subsection{Remark about compression} 
The surjection $\Base\CCC(\Nn) \tto{\gamma^{IM}_I} \CCC(M)$ allows \emph{compressing\/} the data  of type $M$ to programs that output these data when executed on empty input. This means that for every datum $I\tto a M$ there is a program $I\tto {p_a} \Nn$ such that $a = \UK_I^M\circ p_a$. Since we did not introduce a notion of size for the data yet, we do not have a way to say that $p_a$ is a \emph{succinct} way to describe $a$. Indeed the program $p_a$ may say "{\tt print $a$}", and be longer than $a$. Nevertheless, at a future step of this work, this compression mechanism will provide an abstract springboard into \emph{Kolmogorov complexity} \cite{Levin-Zvonkin,Vitanyi:book}, and a foundation for a monoidal view of randomized computation. 

\subsection{Remark about the composition}
The bracket notation provides a convenient way to capture program composition with a minimal structural blowup. The surjectiveness $\{-\} = \gamma^{LM}_I:\Base\CCC(\Nn) \epi  \CCC(L,M)$,  $\{-\} = \gamma^{MN}_I:\Base\CCC(\Nn) \epi  \CCC(M,N)$, and $\{-\}  = \gamma^{LN}_I:\Base\CCC(\Nn) \epi  \CCC(L,N)$ suggests that for any pair of programs $p, q\in \Base\CCC(\Nn)$ such that $\{p\} \in \CCC(L,M)$ and $\{q\}\in \CCC(M,N)$, we can find some $r\in \Base\CCC(\Nn)$ such that $\{r\} = \{q\}\circ \{p\}$. Similarly, for any pair of programs $s,t\in \Base\CCC(\Nn)$ we can find a program $v\in \Base\CCC(\Nn)$ such that $\{v\} = \{s\}\otimes \{t\}$. The type $\Nn$ in $\CCC$ could thus be coherently extended  by partial  monoid operations $-\pcomp-$ and $-\comp-$ such that
\[
\{p\comp q\}\ =\ \{q\} \circ \{p\}  \qquad \qquad \{s\pcomp t\}\ =\ \{s\}\otimes \{t\}
\]
In the standard computer models, this corresponds to introducing the composition operators in the programming language. One could indeed prove that any monoidal computer can be conservatively extended into a monoidal computer where the universal evaluators respect the internal program composition operations, used e.g. in \cite{PavlovicD:NSPW11}. This conservative extension, of course, contains exactly the same information as Def.~\ref{def-moncomp}. Since formalizing it does not seem to offer significant advantages for the present analyses, we shall for the moment keep program composition as syntactic sugar, and 
\begin{itemize}
\item generically denote by $p\comp q$ any program satisfying $\{p\comp q\} = \{q\}\circ \{p\}$, and
\item generically denote by $s\pcomp t$ any program satisfying $\{s\pcomp t\} = \{s\}\otimes  \{t\}$.
\end{itemize}

\section{Arithmetic and logic in monoidal computer}\label{Sec-Arithmetic}
The simplest avenue towards an implementation of the basic logical and recursion theoretic constructions in monoidal computer seems to be by the way of interpreting the monoidal computations as $\lambda$-expressions.

\subsection{Monoidal computer and $\lambda$-calculus}
Viewed from through the lense of $\lambda$-calculus \cite{BarendregtH:book}, the universal evaluators perform the \emph{function application}\/ operation, whereas the partial evaluators perform the \emph{function abstraction}\/ operation
\vspace{-.5\baselineskip}
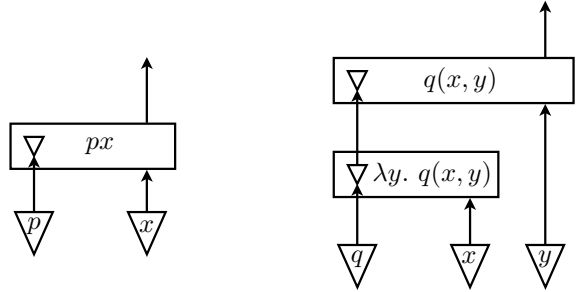
\[\def\JPicScale{1}\renewcommand{\dh}{p}\newcommand{\ahh}{px}\newcommand{\bhh}{x} 
\ifx\JPicScale\undefined\def\JPicScale{1}\fi
\psset{unit=\JPicScale mm}
\psset{linewidth=0.3,dotsep=1,hatchwidth=0.3,hatchsep=1.5,shadowsize=1,dimen=middle}
\psset{dotsize=0.7 2.5,dotscale=1 1,fillcolor=black}
\psset{arrowsize=1 2,arrowlength=1,arrowinset=0.25,tbarsize=0.7 5,bracketlength=0.15,rbracketlength=0.15}
\begin{pspicture}(0,0)(21.88,30.62)
\rput(3.12,8.12){$\dh$}
\psline{->}(3.12,10)(3.12,17.5)
\psline{->}(18.12,10)(18.12,15.62)
\rput(18.12,8.12){$\bhh$}
\pspolygon[](0,21.76)(21.88,21.76)(21.88,15.75)(0,15.75)
\rput(11.88,18.76){$\ahh$}
\psline{->}(18.12,21.88)(18.12,30.62)
\pspolygon(1.88,20)
(4.38,20)
(3.12,17.5)(1.88,20)
\pspolygon(0.62,10)
(5.62,10)
(3.12,4.38)(0.62,10)
\pspolygon(15.62,10)
(20.62,10)
(18.13,4.38)(15.62,10)
\end{pspicture}

\qquad\qquad\qquad
\def\JPicScale{1}\newcommand{\chh}{y}\renewcommand{\dh}{q}\renewcommand{\ahh}{\lambda y.\ q(x,y)}\renewcommand{\bhh}{x}\newcommand{\qhh}{q(x,y)} 
\ifx\JPicScale\undefined\def\JPicScale{1}\fi
\psset{unit=\JPicScale mm}
\psset{linewidth=0.3,dotsep=1,hatchwidth=0.3,hatchsep=1.5,shadowsize=1,dimen=middle}
\psset{dotsize=0.7 2.5,dotscale=1 1,fillcolor=black}
\psset{arrowsize=1 2,arrowlength=1,arrowinset=0.25,tbarsize=0.7 5,bracketlength=0.15,rbracketlength=0.15}
\begin{pspicture}(0,0)(31.88,38.12)
\rput(3.12,3.75){$\dh$}
\psline{->}(3.12,5.62)(3.12,13.75)
\psline{->}(18.12,5.62)(18.12,11.88)
\rput(18,4){$\bhh$}
\pspolygon[](0,18.01)(21.88,18.01)(21.88,12)(0,12)
\rput(13.12,15){$\ahh$}
\psline{->}(28.12,5.62)(28.12,24.38)
\rput(28.12,3.75){$\chh$}
\psline{->}(3.12,16.25)(3.12,26.25)
\pspolygon[](0,30.5)(31.88,30.5)(31.88,24.5)(0,24.5)
\psline{->}(28.12,30.62)(28.12,38.12)
\pspolygon(15.62,5.62)
(20.62,5.62)
(18.12,0)(15.62,5.62)
\pspolygon(0.62,5.62)
(5.62,5.62)
(3.14,0)(0.62,5.62)
\pspolygon(25.62,5.62)
(30.62,5.62)
(28.14,0)(25.62,5.62)
\pspolygon(1.88,28.75)
(4.38,28.75)
(3.12,26.25)(1.88,28.75)
\pspolygon(1.88,16.25)
(4.38,16.25)
(3.12,13.75)(1.88,16.25)
\rput(16.88,27.5){$\qhh$}
\end{pspicture}

\]
\medskip
Together with the data services, these basic operations allow a sound interpretation of any $\lambda$-expression.
\be{prop}\label{prop-lambda}
Every monoidal computer provides a model of nonextensional untyped $\lambda$-calculus.
\ee{prop}

The proof is straightforward, provided that the notion of a model of $\lambda$-calculus is spelled out \cite[I.5.2 and V]{BarendregtH:book}. The terms are interpreted as the elements of the universal data type $\Nn$.  For the terms $t,s\in \CCC(\Nn)$, the abstraction $\lambda x.\ t$ is interpreted by $\SK_{I\Nn}\circ t$, whereas the application $ts$ is $\UK_\Nn^\Nn\circ(t\otimes s)$.  The free variables can be interpreted in the \emph{polynomial extensions\/} of monoidal computers, which are a special case of the constructions in \cite{PavlovicD:MSCS97}. We shall also use the polynomial extensions to capture randomized computation in monoidal computers.

\subsection{Representing numbers and truth values}\label{sec-numbers}
\paragraph{The external view of data computations.} As the data in a monoidal computer are the elements of its data types, it will be convenient to to introduce the notation
\[ \Nnn = \Base\CCC(I,\Nn) \qquad \mbox{ and } \qquad \widehat M = \Base\CCC\left(I, M\right) \]
Note that the data service induces a bijection $\widehat{\Nn^{\otimes m}} \cong \Nnn^m$, so that every computaion, \emph{viz.\/} a morphism in the monoidal computer $\CCC$, induces a \emph{computable function}
\[\prooftree
f\in \CCC\left(\Nn^{\otimes m}, \Nn^{\otimes n}\right)
\justifies
\widehat f \in \Set(\Nnn^m, \Nnn^n)
\endprooftree\]
with $\widehat f(a) = f\circ a$. \emph{Externalizing\/} the computations in this way, from a monidal computer $\CCC$ to the sets of data represented in it, we recover the usual notion of a computable function. 

\be{prop}\label{prop-logic-arithmetic}
Every monoidal computer contains the representations of 
\begin{itemize}
\item truth values $\Bits =\{\tru,\fls\} \hookrightarrow \Nnn$, and
\item natural numbers $\NNn = \{0,1,2,\ldots, n, \ldots\} \hookrightarrow \Nnn$.
\end{itemize}
The computations of the monoidal computer induce just the partial recursive functions on $\NNn$. The logical operations on $\Bits$ are also representable, thus supporting a full model of arithmetic.
\ee{prop}

The representations of logical formulas and of partial recursive functions in untyped $\lambda$-calculus are developed in detail in \cite[Ch.~6]{BarendregtH:book}, or \cite[Ch.~2]{KrivineJL:book}. The original Church's representations
\begin{gather}
\enco\tru \ = \ \lambda px.\ p \notag\\
 \enco\fls\ =\   \enco 0\  =\  \lambda px.\ x \label{eq-totel}\\ 
 \enco{i+1}\ =\ \lambda px.\ p(ipx)  \notag
\end{gather}

%
%
seem the most convenient for our purposes. 
%
%
The basic logical operations are
\beq\label{eq-logic}
\enco{\neg} \ = \lambda xyz.\ xzy \qquad \qquad  \enco{\wedge} \ =\ \lambda xy.\ xyx 
\eeq
whereas the arithmetic operations are constructed using the recursion schema, which can be represented by applying a fixed point operator like the one in Prop.~\ref{prop-fixpoint} to the recursion specifications expressed using the successor $\sigma$ and the $\ifthenelse$ operators
\beq\label{eq-arithmetic}
\enco{\sigma} \ =\ \lambda npx.\ p(npx) \qquad \qquad  \enco{\ifthenelse}\ =\ \lambda bxy.\ bxy 
\eeq
The fact that the $\lambda$-constructions in the monoidal computer do not satisfy the $\eta$-rule $\lambda fx.\ fx = f$ invalidates the uniqueness claims of some constructions, but leaves the representability claims unchanged. Restating all this categorically, in terms of natural numbers object \cite{Lambek-Scott:book,Pare-Roman}, we have

\begin{prop}\label{prop-NNO}
In every monoidal computer, the universal data type $\Nn$ with the structure
$$I \tto{\ 0\ }\Nn\oot {\ \sigma\ } \Nn$$ 
is a weak natural numbers object, in the sense that every pair $L\tto g M \oot h M$ induces a computation $\Nn \otimes L \tto{\cata{g,h}} M$, not necessarily unique, which makes the following diagram commute
\newcommand{\suc}{\sigma}

\beq\label{eq-NNO}
\begin{split}
\def\JPicScale{.8}
\newcommand{\Lhh}{L}
\newcommand{\UpOne}{\Nn\otimes L}
\newcommand{\UpTwo}{\Nn\otimes L}
\newcommand{\DownOne}{M}
\newcommand{\DownTwo}{M}
\newcommand{\one}{\scriptstyle 0\otimes L}
\newcommand{\two}{\scriptstyle g} 
\newcommand{\three}{\scriptstyle \suc\otimes L}
\newcommand{\four}{\scriptstyle h} 
\newcommand{\five}{\scriptstyle \cata{g,h}}
\newcommand{\six}{\scriptstyle \cata{g,h}} 
\ifx\JPicScale\undefined\def\JPicScale{1}\fi
\psset{unit=\JPicScale mm}
\psset{linewidth=0.3,dotsep=1,hatchwidth=0.3,hatchsep=1.5,shadowsize=1,dimen=middle}
\psset{dotsize=0.7 2.5,dotscale=1 1,fillcolor=black}
\psset{arrowsize=1 2,arrowlength=1,arrowinset=0.25,tbarsize=0.7 5,bracketlength=0.15,rbracketlength=0.15}
\begin{pspicture}(0,0)(66.26,32.5)
\rput(5,15){$\Lhh$}
\rput(35,30){$\UpOne$}
\rput(65,30){$\UpTwo$}
\rput(35,0){$\DownOne$}
\psline{->}(8.12,16.88)(28.75,28.75)
\psline{->}(8.75,13.12)(31.88,1.25)
\psline{->}(58.12,30)(42.5,30)
\psline{->}(61.88,0)(37.5,0)
\psline[linestyle=dashed,dash=1 1]{->}(35,26.88)(35,4.38)
\psline[linestyle=dashed,dash=1 1]{->}(65,26.88)(65,4.38)
\rput(65,0){$\DownTwo$}
\rput{25}(16.25,23.75){$\one$}
\rput{-25}(17.5,6.25){$\two$}
\rput(50,32.5){$\three$}
\rput(50,-2.5){$\four$}
\rput[l](36.26,15.62){$\five$}
\rput[l](66.26,15.62){$\six$}
\end{pspicture}
 
\end{split}
\eeq
\end{prop}

\subsection{Numeric monoidal computer}

\paragraph{Notation and terminology.} In a monoidal computer, we use the term \emph{total elements}\/ to refer jointly to the numbers and the truth values, e.g. as represented in Eqn.~\eqref{eq-totel}. The set of all total elements is thus denoted by
\bear
\Totel & = & \Bits \cup \NNn\ \subseteq \ \Nnn
\eear

\be{defn}\label{def-total}
A computation $f\in \CCC\left( \Nn^{\otimes m},\Nn^{\otimes n}\right)$, is \emph{total} if the induced function $\widehat f : \Nnn^m \to \Nnn^n$ maps total elements to total elements, i.e. restricts to a function $\widehat f_{\restriction}:\Totel^m \to \Totel^n$. It is \emph{total numeric} if it maps numbers to numbers, i.e. restricts to a function $\widehat f_{\restriction}:\NNn^m \to \NNn^n$.  Computations that are not total are called \emph{partial}.
\ee{defn}

\be{defn}\label{def-predicate}
A computation $\varphi\in \CCC\left( \Nn^{\otimes m},\Nn\right)$, is a \emph{predicate} if its only total values are $\tru$ or $\fls$, i.e. if $\widehat \varphi \circ a \in \TTt \Longrightarrow  \widehat \varphi \circ a \in \Bits$.
\ee{defn}

\be{defn}\label{def-numeric}
A monoidal computer is \emph{numeric} if
\begin{itemize}
\item every computation has a numeric program: for every
computation $f:L\to M$ there is a number $p\in \NNn$ such that $
f  =   \UK^M_L(p\otimes L)$ 
\item the partial evaluations are total numeric functions $\SK_{LM}: \NNn\otimes L  \to  \NNn$

\item there is a predicate $\downarrow_\NNn$ over $\Nn$ such that 
\bea\label{eq-charnum}
\downarrow_\NNn x & = &  \begin{cases} 
\tru & \mbox{ if } x\in \NNn\\
\fls & \mbox{ otherwise}\end{cases}
\eea 
\end{itemize}
\ee{defn}

\paragraph{Examples and non-examples.} The classical and the quantum computers are numeric, whereas the extensional computer is not.

\paragraph{Restriction.} Henceforth we focus on numeric monoidal computers.

\section{The fixed point constructions} \label{Sec-Fixpoints}
In this section we illustrate the workings of the monoidal computer by spelling out the basic constructions of computability theory.

\begin{prop}\label{prop-fixpoint}
Every computation in a monoidal computer has a fixed point.
\end{prop}

\be{lemma}\label{lemma-Phi}
There is a computable program transformation $\Phi:\Nn\to \Nn$ which executes each program on itself, i.e.
\bear
\Phi\circ p & = & \{p\}\circ p
\eear
\end{lemma}

\bprf{ of Lemma \ref{lemma-Phi}}
The transformation

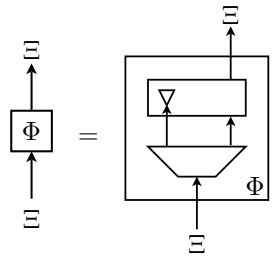
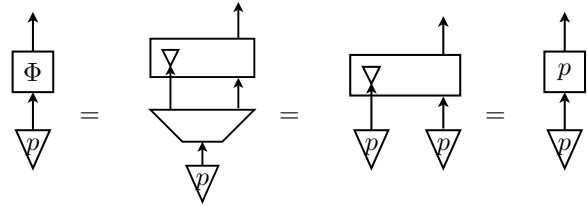
\bear
\def\JPicScale{.9}\newcommand{\aah}{\Nn}\renewcommand{\dh}{\Nn}\newcommand{\ahh}{\Phi} 
\ifx\JPicScale\undefined\def\JPicScale{1}\fi
\psset{unit=\JPicScale mm}
\psset{linewidth=0.3,dotsep=1,hatchwidth=0.3,hatchsep=1.5,shadowsize=1,dimen=middle}
\psset{dotsize=0.7 2.5,dotscale=1 1,fillcolor=black}
\psset{arrowsize=1 2,arrowlength=1,arrowinset=0.25,tbarsize=0.7 5,bracketlength=0.15,rbracketlength=0.15}
\begin{pspicture}(0,0)(8,13.75)
\rput(5,13.75){$\dh$}
\psline{->}(5,-8.25)(5,-1.25)
\pspolygon[](2,4.75)(8,4.75)(8,-1.25)(2,-1.25)
\psline{->}(5,4.75)(5,11.75)
\rput(5,1.75){$\ahh$}
\rput(5,-11.25){$\aah$}
\end{pspicture}
 & = & \def\JPicScale{.8}\newcommand{\aah}{\Phi}\renewcommand{\dh}{\Nn}\newcommand{\ahh}{} \newcommand{\diagonal}{}
\ifx\JPicScale\undefined\def\JPicScale{1}\fi
\psset{unit=\JPicScale mm}
\psset{linewidth=0.3,dotsep=1,hatchwidth=0.3,hatchsep=1.5,shadowsize=1,dimen=middle}
\psset{dotsize=0.7 2.5,dotscale=1 1,fillcolor=black}
\psset{arrowsize=1 2,arrowlength=1,arrowinset=0.25,tbarsize=0.7 5,bracketlength=0.15,rbracketlength=0.15}
\begin{pspicture}(0,0)(23.75,21.25)
\rput(17.5,21.25){$\dh$}
\psline{->}(6.88,-0.62)(6.88,6.38)
\pspolygon[](3.75,10.5)(20,10.5)(20,4.5)(3.75,4.5)
\psline{->}(17.5,10.62)(17.5,19.38)
\rput(17.5,7.5){$\ahh$}
\rput[Br](23.12,-8.75){$\aah$}
\psline{->}(17.5,-0.38)(17.5,4.38)
\pspolygon(3.75,-0.62)
(20,-0.62)
(15,-5.62)
(8.75,-5.62)(3.75,-0.62)
\psline{->}(11.88,-14.38)(11.88,-5.62)
\rput(11.88,-3.12){$\diagonal$}
\pspolygon[](0,14.38)(23.75,14.38)(23.75,-9.5)(0,-9.5)
\rput(11.88,-16.88){$\dh$}
\pspolygon(5.62,8.75)
(8.12,8.75)
(6.88,6.25)(5.62,8.75)
\end{pspicture}

\eear

\bigskip
\bigskip

satisfies

\bear
\def\JPicScale{.9}\newcommand{\ahh}{\Phi}\newcommand{\bhh}{p}  
\ifx\JPicScale\undefined\def\JPicScale{1}\fi
\psset{unit=\JPicScale mm}
\psset{linewidth=0.3,dotsep=1,hatchwidth=0.3,hatchsep=1.5,shadowsize=1,dimen=middle}
\psset{dotsize=0.7 2.5,dotscale=1 1,fillcolor=black}
\psset{arrowsize=1 2,arrowlength=1,arrowinset=0.25,tbarsize=0.7 5,bracketlength=0.15,rbracketlength=0.15}
\begin{pspicture}(0,0)(8,16.25)
\psline{->}(5,-1.25)(5,4.38)
\pspolygon[](2,10.38)(8,10.38)(8,4.38)(2,4.38)
\psline{->}(5,10.38)(5,16.25)
\rput(5,7.38){$\ahh$}
\rput(5,-3.75){$\bhh$}
\pspolygon(2.5,-1.25)
(7.5,-1.25)
(5,-6.88)(2.5,-1.25)
\end{pspicture}
 & = & \def\JPicScale{.85}\newcommand{\bhh}{p}\newcommand{\aah}{\Phi}\renewcommand{\dh}{\Nn}\newcommand{\ahh}{} \newcommand{\diagonal}{}
\ifx\JPicScale\undefined\def\JPicScale{1}\fi
\psset{unit=\JPicScale mm}
\psset{linewidth=0.3,dotsep=1,hatchwidth=0.3,hatchsep=1.5,shadowsize=1,dimen=middle}
\psset{dotsize=0.7 2.5,dotscale=1 1,fillcolor=black}
\psset{arrowsize=1 2,arrowlength=1,arrowinset=0.25,tbarsize=0.7 5,bracketlength=0.15,rbracketlength=0.15}
\begin{pspicture}(0,0)(20,18.75)
\psline{->}(6.88,1.88)(6.88,8.88)
\pspolygon[](3.75,13)(20,13)(20,7)(3.75,7)
\psline{->}(17.5,13.12)(17.5,18.75)
\rput(17.5,10){$\ahh$}
\psline{->}(17.5,2.12)(17.5,6.88)
\pspolygon(3.75,1.88)
(20,1.88)
(15,-3.12)
(8.75,-3.12)(3.75,1.88)
\psline{->}(11.88,-6.88)(11.88,-3.12)
\rput(11.88,-0.62){$\diagonal$}
\rput(11.88,-9.38){$\bhh$}
\pspolygon(5.62,11.25)
(8.12,11.25)
(6.88,8.75)(5.62,11.25)
\pspolygon(9.38,-6.88)
(14.38,-6.88)
(11.88,-12.5)(9.38,-6.88)
\end{pspicture}
\ \ =\ \ \def\JPicScale{.9}
\ifx\JPicScale\undefined\def\JPicScale{1}\fi
\psset{unit=\JPicScale mm}
\psset{linewidth=0.3,dotsep=1,hatchwidth=0.3,hatchsep=1.5,shadowsize=1,dimen=middle}
\psset{dotsize=0.7 2.5,dotscale=1 1,fillcolor=black}
\psset{arrowsize=1 2,arrowlength=1,arrowinset=0.25,tbarsize=0.7 5,bracketlength=0.15,rbracketlength=0.15}
\begin{pspicture}(0,0)(20,15.62)
\psline{->}(6.88,-1.25)(6.88,5.75)
\pspolygon[](3.75,9.88)(20,9.88)(20,3.88)(3.75,3.88)
\psline{->}(17.5,10)(17.5,15.62)
\rput(17.5,6.88){$\ahh$}
\psline{->}(17.5,-1)(17.5,3.75)
\rput(6.88,-3.75){$\bhh$}
\pspolygon(4.38,-1.25)
(9.38,-1.25)
(6.88,-6.88)(4.38,-1.25)
\rput(17.5,-3.75){$\bhh$}
\pspolygon(5.62,8.12)
(8.12,8.12)
(6.88,5.62)(5.62,8.12)
\pspolygon(15,-1.25)
(20,-1.25)
(17.5,-6.88)(15,-1.25)
\end{pspicture}
 \ \ =\ \ \def\JPicScale{.9}\renewcommand{\ahh}{p}  
\ifx\JPicScale\undefined\def\JPicScale{1}\fi
\psset{unit=\JPicScale mm}
\psset{linewidth=0.3,dotsep=1,hatchwidth=0.3,hatchsep=1.5,shadowsize=1,dimen=middle}
\psset{dotsize=0.7 2.5,dotscale=1 1,fillcolor=black}
\psset{arrowsize=1 2,arrowlength=1,arrowinset=0.25,tbarsize=0.7 5,bracketlength=0.15,rbracketlength=0.15}
\begin{pspicture}(0,0)(8,16.25)
\psline{->}(5,-1.25)(5,4.38)
\pspolygon[](2,10.38)(8,10.38)(8,4.38)(2,4.38)
\psline{->}(5,10.38)(5,16.25)
\rput(5,7.38){$\ahh$}
\rput(5,-3.75){$\bhh$}
\pspolygon(2.5,-1.25)
(7.5,-1.25)
(5,-6.88)(2.5,-1.25)
\end{pspicture}

\eear

\vspace{2\baselineskip}
because the programs $p$ are basic data, and thus satisfy $\cmn\circ p = p\otimes p$.
\epr

\bigskip

\paragraph{Remark.} Note that the second step of the above proof essentially depends on the stipulation in Def.~\ref{def-moncomp}, that programs are \emph{basic} data, and thus copiable by the data service.

\bprf{ of Prop.~\ref{prop-fixpoint}} The fixed point of an arbitrary $p\in \CCC(\Nn)$ is
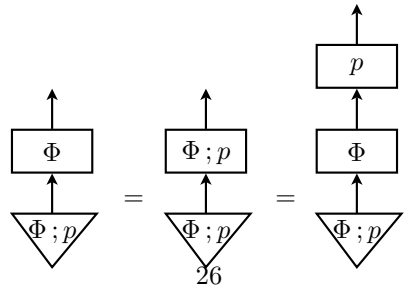
\bear
\def\JPicScale{.95}\newcommand{\ahh}{\Phi}\newcommand{\bhh}{\Phi \comp p}  
\ifx\JPicScale\undefined\def\JPicScale{1}\fi
\psset{unit=\JPicScale mm}
\psset{linewidth=0.3,dotsep=1,hatchwidth=0.3,hatchsep=1.5,shadowsize=1,dimen=middle}
\psset{dotsize=0.7 2.5,dotscale=1 1,fillcolor=black}
\psset{arrowsize=1 2,arrowlength=1,arrowinset=0.25,tbarsize=0.7 5,bracketlength=0.15,rbracketlength=0.15}
\begin{pspicture}(0,0)(11.25,16.25)
\psline{->}(5,-1.25)(5,4.38)
\pspolygon[](-0.62,10.5)(10.62,10.5)(10.62,4.5)(-0.62,4.5)
\psline{->}(5,10.62)(5,16.25)
\rput(5,7.38){$\ahh$}
\rput(5,-3.75){$\bhh$}
\pspolygon(-0.62,-1.25)
(11.25,-1.25)
(5,-8.75)(-0.62,-1.25)
\end{pspicture}
 & = & \def\JPicScale{.95}\newcommand{\ahh}{\Phi\comp p}\newcommand{\bhh}{\Phi \comp p}  
\ifx\JPicScale\undefined\def\JPicScale{1}\fi
\psset{unit=\JPicScale mm}
\psset{linewidth=0.3,dotsep=1,hatchwidth=0.3,hatchsep=1.5,shadowsize=1,dimen=middle}
\psset{dotsize=0.7 2.5,dotscale=1 1,fillcolor=black}
\psset{arrowsize=1 2,arrowlength=1,arrowinset=0.25,tbarsize=0.7 5,bracketlength=0.15,rbracketlength=0.15}
\begin{pspicture}(0,0)(11.25,16.25)
\psline{->}(5,-1.25)(5,4.38)
\pspolygon[](-0.62,10.5)(10.62,10.5)(10.62,4.5)(-0.62,4.5)
\psline{->}(5,10.62)(5,16.25)
\rput(5,7.38){$\ahh$}
\rput(5,-3.75){$\bhh$}
\pspolygon(-0.62,-1.25)
(11.25,-1.25)
(5,-8.75)(-0.62,-1.25)
\end{pspicture}
 \ \ =\ \ \def\JPicScale{.95}\renewcommand{\ahh}{\Phi}\newcommand{\chh}{p}  
\ifx\JPicScale\undefined\def\JPicScale{1}\fi
\psset{unit=\JPicScale mm}
\psset{linewidth=0.3,dotsep=1,hatchwidth=0.3,hatchsep=1.5,shadowsize=1,dimen=middle}
\psset{dotsize=0.7 2.5,dotscale=1 1,fillcolor=black}
\psset{arrowsize=1 2,arrowlength=1,arrowinset=0.25,tbarsize=0.7 5,bracketlength=0.15,rbracketlength=0.15}
\begin{pspicture}(0,0)(11.25,28.12)
\psline{->}(5,-1.25)(5,4.38)
\pspolygon[](-0.62,10.5)(10.62,10.5)(10.62,4.5)(-0.62,4.5)
\psline{->}(5,10.62)(5,16.25)
\rput(5,7.38){$\ahh$}
\rput(5,-3.75){$\bhh$}
\pspolygon(-0.62,-1.25)
(11.25,-1.25)
(5,-8.75)(-0.62,-1.25)
\pspolygon[](-0.62,22.38)(10.62,22.38)(10.62,16.38)(-0.62,16.38)
\psline{->}(5,22.5)(5,28.12)
\rput(5,19.38){$\chh$}
\end{pspicture}

\eear

\bigskip
\bigskip
where $\Phi\comp p$ denotes any program such that $\{\Phi\comp p\} = \{p\}\circ \{\Phi\}$.
\eprf

\begin{corollary}
Every nontrivial numeric computer contains partial computations. Its universal data type always contains non-numeric values.
\end{corollary}

\bpr
Consider the successor operation $\sigma =  \lambda npx.\ p(npx)$ and its fixed point $\bot = \{\Phi\}(\Phi\comp \sigma)$. Since the soundness of Church's numeral representation \cite[Sec.~6.4]{BarendregtH:book} implies that $\sigma n\neq n$ holds for all $n\in \NNn$, it follows that $\bot \not \in \NNn$. On the other hand, since the computer is assumed to be numeric, we can choose the program $\Phi\comp \sigma\in \NNn$. Since  $\{\Phi\}(\Phi\comp \sigma) = \bot$, the computation $\{\Phi\}$ maps an element of $\NNn$ outside $\NNn$, and is therefore not total.
\epr

\section{Kleene's Second  Recursion Theorem}\label{Sec-Kleene}
This theorem is undoubtedly one of the stepping stones into computability theory \cite{MoschovakisY:Kleene}. It says that every total computation $\widehat t:\NNn\to \NNn$, viewed as a program transformation, has a \emph{fixed program}\footnote{Note that there are many total functions which do not have any \emph{fixed numbers}: e.g. the successor function satisfies $\sigma(n) \neq n$ for all $n\in \NNn$.} $p_t\in \NNn$, which encodes the same computable function as its $t$-image, i.e.
\bear
\{p_t\} & = & \left\{t(p_t)\right\}
\eear
\begin{prop}\label{prop-kleene}
In any numeric monoidal computer $\CCC$, for every computation $t:\Nn\to \Nn$, which induces a total function $\widehat t: \NNn\to \NNn$, and for any two types $L,M\in |\CCC|$ there is a program $p_t \in \NNn$ which evaluates to the same $L\to M$ computation like $t\circ p_t$, i.e.
\bear
\UK^M_L \circ(p \otimes L) & = & \UK^M_L\circ (tp_t \otimes L)
\eear 
\end{prop}

\paragraph{Proof.}
Define
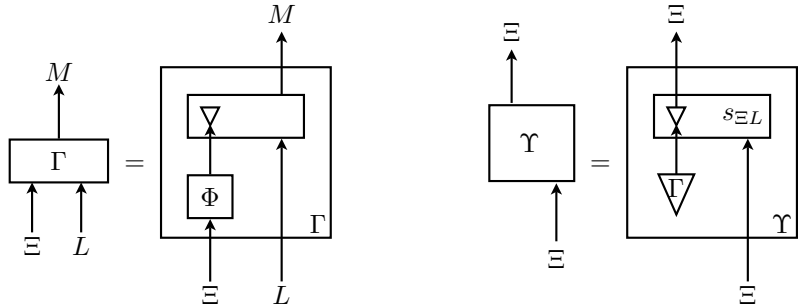
\[
\def\JPicScale{.95}\newcommand{\eh}{L}\newcommand{\fh}{M}\renewcommand{\dh}{\Nn}\newcommand{\ahh}{\Gamma} 
\ifx\JPicScale\undefined\def\JPicScale{1}\fi
\psset{unit=\JPicScale mm}
\psset{linewidth=0.3,dotsep=1,hatchwidth=0.3,hatchsep=1.5,shadowsize=1,dimen=middle}
\psset{dotsize=0.7 2.5,dotscale=1 1,fillcolor=black}
\psset{arrowsize=1 2,arrowlength=1,arrowinset=0.25,tbarsize=0.7 5,bracketlength=0.15,rbracketlength=0.15}
\begin{pspicture}(0,0)(11.88,13.75)
\rput(1.25,-10.62){$\dh$}
\psline{->}(1.25,-8.75)(1.25,-1.75)
\pspolygon[](-1.88,4.25)(11.88,4.25)(11.88,-1.75)(-1.88,-1.75)
\psline{->}(5,4.38)(5,12)
\rput(5,1.25){$\ahh$}
\psline{->}(8.12,-8.75)(8.12,-1.75)
\rput(8.12,-10.62){$\eh$}
\rput(5,13.75){$\fh$}
\end{pspicture}
\ = \ \def\JPicScale{.95}\newcommand{\aah}{\Gamma}\renewcommand{\dh}{\Nn}\renewcommand{\ahh}{} \newcommand{\diagonal}{\Phi}\renewcommand{\eh}{L}\renewcommand{\fh}{M}
\ifx\JPicScale\undefined\def\JPicScale{1}\fi
\psset{unit=\JPicScale mm}
\psset{linewidth=0.3,dotsep=1,hatchwidth=0.3,hatchsep=1.5,shadowsize=1,dimen=middle}
\psset{dotsize=0.7 2.5,dotscale=1 1,fillcolor=black}
\psset{arrowsize=1 2,arrowlength=1,arrowinset=0.25,tbarsize=0.7 5,bracketlength=0.15,rbracketlength=0.15}
\begin{pspicture}(0,0)(23.75,21.88)
\psline{->}(6.88,-0.62)(6.88,6.38)
\pspolygon[](3.75,10.5)(20,10.5)(20,4.5)(3.75,4.5)
\psline{->}(16.88,10.62)(16.88,19.38)
\rput(16.88,7.5){$\ahh$}
\rput[Br](23.12,-8.75){$\aah$}
\psline{->}(16.88,-15.62)(16.88,4.38)
\psline{->}(6.88,-15.64)(6.88,-6.88)
\rput(6.88,-3.75){$\diagonal$}
\pspolygon[](0,14.38)(23.75,14.38)(23.75,-9.5)(0,-9.5)
\rput(6.88,-17.5){$\dh$}
\pspolygon[](3.75,-0.75)(10,-0.75)(10,-6.75)(3.75,-6.75)
\rput(16.88,-17.5){$\eh$}
\rput(16.88,21.88){$\fh$}
\pspolygon(5.62,8.75)
(8.12,8.75)
(6.88,6.25)(5.62,8.75)
\end{pspicture}
 
\qquad\qquad \qquad
\def\JPicScale{.95}\newcommand{\HaH}{\Upsilon}\newcommand{\chh}{\Nn} 
\ifx\JPicScale\undefined\def\JPicScale{1}\fi
\psset{unit=\JPicScale mm}
\psset{linewidth=0.3,dotsep=1,hatchwidth=0.3,hatchsep=1.5,shadowsize=1,dimen=middle}
\psset{dotsize=0.7 2.5,dotscale=1 1,fillcolor=black}
\psset{arrowsize=1 2,arrowlength=1,arrowinset=0.25,tbarsize=0.7 5,bracketlength=0.15,rbracketlength=0.15}
\begin{pspicture}(0,0)(11.88,19.38)
\pspolygon[](0,9.06)(11.88,9.06)(11.88,-1.56)(0,-1.56)
\psline{->}(3.12,9.38)(3.12,16.88)
\psline{->}(9.38,-10)(9.38,-1.87)
\rput(5.62,3.75){$\HaH$}
\rput(9.38,-12.5){$\chh$}
\rput(3.12,19.38){$\chh$}
\end{pspicture}
 \ = \ \def\JPicScale{.95}\renewcommand{\aah}{\Upsilon}\renewcommand{\ahh}{\SK_{\Nn L}} \renewcommand{\eh}{\Nn}\renewcommand{\fh}{\Nn}\newcommand{\bhh}{\Gamma}
\ifx\JPicScale\undefined\def\JPicScale{1}\fi
\psset{unit=\JPicScale mm}
\psset{linewidth=0.3,dotsep=1,hatchwidth=0.3,hatchsep=1.5,shadowsize=1,dimen=middle}
\psset{dotsize=0.7 2.5,dotscale=1 1,fillcolor=black}
\psset{arrowsize=1 2,arrowlength=1,arrowinset=0.25,tbarsize=0.7 5,bracketlength=0.15,rbracketlength=0.15}
\begin{pspicture}(0,0)(23.75,21.88)
\psline{->}(6.88,-0.62)(6.88,6.38)
\pspolygon[](3.75,10.5)(20,10.5)(20,4.5)(3.75,4.5)
\psline{->}(6.88,8.75)(6.88,19.38)
\rput(16.25,7.5){$\ahh$}
\rput[Br](23.12,-8.75){$\aah$}
\psline{->}(16.88,-15.62)(16.88,4.38)
\pspolygon[](0,14.38)(23.75,14.38)(23.75,-9.5)(0,-9.5)
\rput(16.88,-17.5){$\eh$}
\rput(6.88,21.88){$\fh$}
\rput(6.88,-2.5){$\bhh$}
\pspolygon(5.62,8.75)
(8.12,8.75)
(6.88,6.25)(5.62,8.75)
\pspolygon(4.38,-0.62)
(9.38,-0.62)
(6.88,-6.25)(4.38,-0.62)
\end{pspicture}

\]

\vspace{5\baselineskip}
The program transformations showing that the program $p_t = \{\Upsilon\}(\Upsilon\comp t)$ is the claimed fixed point are displayed on the following diagram

%

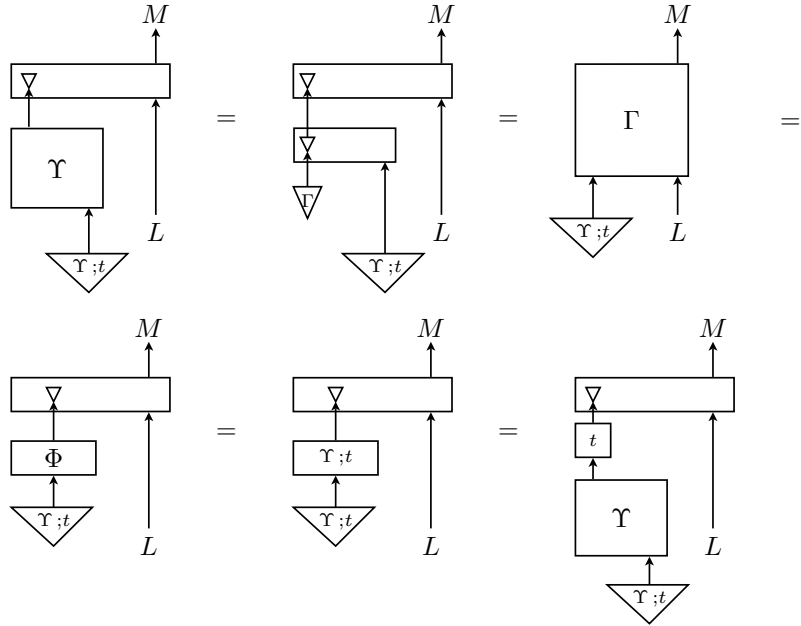
\[\def\JPicScale{.75}\renewcommand{\dh}{\Nn}\newcommand{\ahh}{}\newcommand{\shh}{}\newcommand{\bhh}{\scriptstyle\Gamma}\newcommand{\Bhh}{\Gamma}\newcommand{\chh}{L}\newcommand{\Lhh}{M}\newcommand{\Ht}{\scriptstyle \Upsilon \comp t}\newcommand{\HaH}{\Upsilon}\newcommand{\FaH}{\Phi} \newcommand{\eqls}{=} \newcommand{\tttt}{\scriptstyle t}
\ifx\JPicScale\undefined\def\JPicScale{1}\fi
\psset{unit=\JPicScale mm}
\psset{linewidth=0.3,dotsep=1,hatchwidth=0.3,hatchsep=1.5,shadowsize=1,dimen=middle}
\psset{dotsize=0.7 2.5,dotscale=1 1,fillcolor=black}
\psset{arrowsize=1 2,arrowlength=1,arrowinset=0.25,tbarsize=0.7 5,bracketlength=0.15,rbracketlength=0.15}
\begin{pspicture}(0,0)(138.12,88.75)
\pspolygon[](50,24.24)(78.12,24.24)(78.12,18.24)(50,18.24)
\rput(73.12,21.24){$\ahh$}
\psline{->}(74.38,-2.49)(74.38,18.12)
\rput(74.38,-5.62){$\chh$}
\psline{->}(74.38,24.38)(74.38,30.62)
\rput(74.38,33.12){$\Lhh$}
\psline{->}(52.5,58.12)(52.5,64.38)
\pspolygon[](50.12,68.5)(68.12,68.5)(68.12,62.5)(50.12,62.5)
\psline{->}(52.5,66.88)(52.5,75.62)
\psline{->}(66.25,46.25)(66.25,62.5)
\pspolygon[](50,79.87)(78.12,79.87)(78.12,73.87)(50,73.87)
\rput(73.12,76.87){$\ahh$}
\rput(65,65.63){$\shh$}
\psline{->}(76.25,53.13)(76.25,73.75)
\rput(76.25,50){$\chh$}
\psline{->}(76.25,80)(76.25,86.25)
\rput(76.25,88.75){$\Lhh$}
\pspolygon[](0,68.44)(16.25,68.44)(16.25,54.38)(0,54.38)
\psline{->}(3.12,68.75)(3.12,75.62)
\pspolygon[](0,79.87)(28.12,79.87)(28.12,73.87)(0,73.87)
\rput(23.12,76.87){$\ahh$}
\psline{->}(25.62,53.13)(25.62,73.75)
\rput(25.62,50){$\chh$}
\psline{->}(25.62,80)(25.62,86.25)
\rput(25.62,88.75){$\Lhh$}
\pspolygon[](100,79.87)(120,79.88)(120,60)(100,59.99)
\psline{->}(118.12,53.13)(118.12,60)
\rput(118.12,50){$\chh$}
\psline{->}(118.12,80)(118.12,86.25)
\rput(118.12,88.75){$\Lhh$}
\pspolygon[](0,13)(15,13)(15,7)(0,7)
\psline{->}(7.5,13.12)(7.5,20)
\pspolygon[](0,24.25)(28.12,24.25)(28.12,18.25)(0,18.25)
\rput(23.12,21.25){$\ahh$}
\psline{->}(24.38,-2.49)(24.38,18.12)
\rput(24.38,-5.62){$\chh$}
\psline{->}(24.38,24.38)(24.38,30.62)
\rput(24.38,33.12){$\Lhh$}
\pspolygon[](100,16.12)(106.25,16.12)(106.25,10.12)(100,10.12)
\psline{->}(103.12,16.25)(103.12,20)
\pspolygon[](100,24.24)(128.12,24.24)(128.12,18.24)(100,18.24)
\rput(123.12,21.24){$\ahh$}
\psline{->}(124.38,-2.49)(124.38,18.12)
\rput(124.38,-5.62){$\chh$}
\psline{->}(124.38,24.38)(124.38,30.62)
\rput(124.38,33.12){$\Lhh$}
\rput(38.12,70){$\eqls$}
\rput(88.12,70){$\eqls$}
\rput(138.12,69.38){$\eqls$}
\rput(88.12,14.38){$\eqls$}
\rput(38.12,14.38){$\eqls$}
\rput(52.5,55.62){$\bhh$}
\pspolygon(58.75,46.25)
(73.12,46.25)
(66.25,39.38)(58.75,46.25)
\rput(66.25,43.75){$\Ht$}
\rput(8.12,61.25){$\HaH$}
\psline{->}(13.75,46.25)(13.75,54.38)
\pspolygon(6.25,46.25)
(20.62,46.25)
(13.75,39.38)(6.25,46.25)
\rput(13.75,43.75){$\Ht$}
\psline{->}(103.12,52.5)(103.12,60)
\pspolygon(95.62,52.5)
(110,52.5)
(103.12,45.62)(95.62,52.5)
\rput(103.12,50){$\Ht$}
\rput(110,70){$\Bhh$}
\psline{->}(7.5,1.25)(7.5,6.88)
\pspolygon(0,1.25)
(14.38,1.25)
(7.5,-5.62)(0,1.25)
\rput(7.5,-1.25){$\Ht$}
\pspolygon[](50,13)(65,13)(65,7)(50,7)
\psline{->}(57.5,13.12)(57.5,20)
\psline{->}(57.5,1.25)(57.5,6.88)
\pspolygon(50,1.25)
(64.38,1.25)
(57.5,-5.62)(50,1.25)
\rput(57.5,-1.25){$\Ht$}
\rput(7.5,10){$\FaH$}
\rput(57.5,10){$\Ht$}
\pspolygon[](100,6.09)(116.25,6.09)(116.25,-7.34)(100,-7.34)
\psline{->}(103.12,6.25)(103.12,10)
\rput(108.12,-0.62){$\HaH$}
\psline{->}(113.12,-12.5)(113.12,-7.5)
\pspolygon(105.62,-12.5)
(120,-12.5)
(113.12,-19.38)(105.62,-12.5)
\rput(113.12,-15){$\Ht$}
\rput(103.12,13.12){$\tttt$}
\pspolygon(6.25,22.5)
(8.75,22.5)
(7.5,20)(6.25,22.5)
\pspolygon(56.25,22.5)
(58.75,22.5)
(57.5,20)(56.25,22.5)
\pspolygon(101.88,22.5)
(104.38,22.5)
(103.12,20)(101.88,22.5)
\pspolygon(51.25,66.88)
(53.75,66.88)
(52.5,64.38)(51.25,66.88)
\pspolygon(51.25,78.12)
(53.75,78.12)
(52.5,75.62)(51.25,78.12)
\pspolygon(1.88,78.12)
(4.38,78.12)
(3.12,75.62)(1.88,78.12)
\pspolygon(50,58.12)
(55,58.12)
(52.51,52.5)(50,58.12)
\end{pspicture}

\]


\vspace{5\baselineskip}
\be{corollary}\label{corollary-kleene}
In any numeric monoidal computer $\CCC$, for every computation $f:\Nn\otimes L\to M$ has a fixed program $p_f \in \NNn$ which evaluates to the same $L\to M$ computation like the partial evaluation of $f$ on it
\bear\{p_f\} & = & \lambda x.\ f(p_f,x)
\eear
or diagrammatically
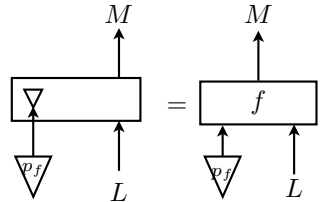
\bear
 \def\JPicScale{.95}\newcommand{\aah}{\scriptstyle p_f}\renewcommand{\dh}{M}\newcommand{\ahh}{}\newcommand{\bhh}{L}  & = & \def\JPicScale{.95}\newcommand{\aah}{\scriptstyle p_f}\newcommand{\eh}{L}\newcommand{\fh}{M}\renewcommand{\dh}{\Nn}\newcommand{\ahh}{f} 
\ifx\JPicScale\undefined\def\JPicScale{1}\fi
\psset{unit=\JPicScale mm}
\psset{linewidth=0.3,dotsep=1,hatchwidth=0.3,hatchsep=1.5,shadowsize=1,dimen=middle}
\psset{dotsize=0.7 2.5,dotscale=1 1,fillcolor=black}
\psset{arrowsize=1 2,arrowlength=1,arrowinset=0.25,tbarsize=0.7 5,bracketlength=0.15,rbracketlength=0.15}
\begin{pspicture}(0,0)(14.38,13.75)
\pspolygon[](-1.88,4.25)(14.38,4.25)(14.38,-1.75)(-1.88,-1.75)
\psline{->}(6.25,4.38)(6.25,11.38)
\rput(6.25,1.25){$\ahh$}
\psline{->}(11.25,-8.75)(11.25,-1.75)
\rput(11.25,-10.62){$\eh$}
\rput(6.25,13.75){$\fh$}
\psline{->}(1.25,-6.25)(1.25,-1.88)
\rput(1.25,-8.75){$\aah$}
\pspolygon(-1.25,-6.25)
(3.75,-6.25)
(1.26,-11.88)(-1.25,-6.25)
\end{pspicture}

\eear 
\ee{corollary}
\vspace{5\baselineskip}

\bpr
Apply Prop~\ref{prop-kleene} to the total computation
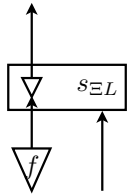
\[
\def\JPicScale{1}\newcommand{\aah}{f}\newcommand{\ahh}{\SK_{\Nn L}}\renewcommand{\dh}{}
\ifx\JPicScale\undefined\def\JPicScale{1}\fi
\psset{unit=\JPicScale mm}
\psset{linewidth=0.3,dotsep=1,hatchwidth=0.3,hatchsep=1.5,shadowsize=1,dimen=middle}
\psset{dotsize=0.7 2.5,dotscale=1 1,fillcolor=black}
\psset{arrowsize=1 2,arrowlength=1,arrowinset=0.25,tbarsize=0.7 5,bracketlength=0.15,rbracketlength=0.15}
\begin{pspicture}(0,0)(17.5,13.12)
\psline{->}(5,-6.25)(5,0.75)
\pspolygon[](1.88,4.88)(17.5,4.88)(17.5,-1.12)(1.88,-1.12)
\psline{->}(5,3.12)(5,13.12)
\rput(13.75,1.88){$\ahh$}
\rput(5,-8.75){$\aah$}
\psline{->}(14.38,-11.88)(14.38,-1.25)
\rput[l](15.62,-4.38){$\dh$}
\rput[r](3.75,8.12){$\dh$}
\rput[r](3.75,-3.75){$\dh$}
\pspolygon(3.75,3.12)
(6.25,3.12)
(5,0.62)(3.75,3.12)
\pspolygon(2.49,-6.25)
(7.49,-6.25)
(5,-11.88)(2.49,-6.25)
\end{pspicture}

\]

\epr

\section{Undecidability of the Halting Problem}\label{Sec-Halting}
\be{defn}
A program is said to \emph{halt} on $n\in \NNn$ if $\{p\}\circ n \in \NNn$. The \emph{halting predicate} is thus defined
\bea\label{eq-halting}
\def\JPicScale{.9}\newcommand{\eh}{\Nn}\newcommand{\fh}{\Nn}\renewcommand{\dh}{\Nn}\newcommand{\ahh}{H} 
\ifx\JPicScale\undefined\def\JPicScale{1}\fi
\psset{unit=\JPicScale mm}
\psset{linewidth=0.3,dotsep=1,hatchwidth=0.3,hatchsep=1.5,shadowsize=1,dimen=middle}
\psset{dotsize=0.7 2.5,dotscale=1 1,fillcolor=black}
\psset{arrowsize=1 2,arrowlength=1,arrowinset=0.25,tbarsize=0.7 5,bracketlength=0.15,rbracketlength=0.15}
\begin{pspicture}(0,0)(11.88,13.75)
\rput(1.25,-10.62){$\dh$}
\psline{->}(1.25,-8.75)(1.25,-1.75)
\pspolygon[](-1.88,4.25)(11.88,4.25)(11.88,-1.75)(-1.88,-1.75)
\psline{->}(8.12,4.38)(8.12,12)
\rput(5,1.25){$\ahh$}
\psline{->}(8.12,-8.75)(8.12,-1.75)
\rput(8.12,-10.62){$\eh$}
\rput(8.12,13.75){$\fh$}
\end{pspicture}
 & = & \def\JPicScale{.8}\newcommand{\aah}{\scriptstyle \Nn}\renewcommand{\dh}{\Nn}\newcommand{\ahh}{} \newcommand{\diagonal}{\downarrow}\newcommand{\eh}{\Nn}\newcommand{\fh}{\Nn}
\ifx\JPicScale\undefined\def\JPicScale{1}\fi
\psset{unit=\JPicScale mm}
\psset{linewidth=0.3,dotsep=1,hatchwidth=0.3,hatchsep=1.5,shadowsize=1,dimen=middle}
\psset{dotsize=0.7 2.5,dotscale=1 1,fillcolor=black}
\psset{arrowsize=1 2,arrowlength=1,arrowinset=0.25,tbarsize=0.7 5,bracketlength=0.15,rbracketlength=0.15}
\begin{pspicture}(0,0)(23.75,22.5)
\psline{->}(6.88,-14.38)(6.88,-4.24)
\pspolygon[](3.75,-0.12)(20,-0.12)(20,-6.12)(3.75,-6.12)
\psline{->}(16.88,11.87)(16.88,20)
\rput(16.88,-3.12){$\ahh$}
\rput[l](17.5,2.5){$\aah$}
\psline{->}(16.88,-14.38)(16.88,-6.25)
\psline{->}(16.88,0)(16.88,5.62)
\rput(16.88,8.75){$\diagonal$}
\pspolygon[](0,14.38)(23.75,14.38)(23.75,-9.5)(0,-9.5)
\rput(6.88,-17.5){$\dh$}
\pspolygon[](13.75,11.75)(20,11.75)(20,5.75)(13.75,5.75)
\rput(16.88,-17.5){$\eh$}
\rput(16.88,22.5){$\fh$}
\pspolygon(5.62,-1.88)
(8.12,-1.88)
(6.88,-4.38)(5.62,-1.88)
\end{pspicture}

\eea

\vspace{3\baselineskip}
using the predicate $\downarrow:\Nn\to \Nn$ from Def.~\ref{def-numeric}.
\ee{defn}

\be{defn}
A predicate $\varphi:M\to \Nn$ is said to be \emph{decidable}\/ if it is total.
\ee{defn}

\paragraph{Explanation.} Recall from Def.~\ref{def-predicate} that a computation $\varphi\in \CCC(M,\Nn)$ is a predicate if $\widehat \varphi\circ a \in \Totel \Longrightarrow \widehat \varphi\circ a \in \Bits$. Recall from Def.~\ref{def-total} that $\varphi$ is total if $a\in \Totel^m \Longrightarrow \widehat \varphi \circ  a \in \Totel$.  A decidable (i.e. total) predicate $\varphi$ thus satisfies $a\in \Totel^m \Longrightarrow \widehat \varphi\circ a \in \Bits$. 

\be{prop}\label{prop-halting}
The halting predicate $H$, defined by Eq.~\eqref{eq-halting}, is undecidable.
\ee{prop}

\be{defn}\label{def-nontrivial}
A predicate $\varphi:M\to \Nn$ is \emph{nontrivial} if there is a computation $\widetilde \varphi:L\to L$ such that $\neg \varphi = \varphi \widetilde \varphi$
\bea\label{eq-tilde}
\def\JPicScale{.9}\newcommand{\ahh}{\varphi}\newcommand{\chh}{\neg}  
\ifx\JPicScale\undefined\def\JPicScale{1}\fi
\psset{unit=\JPicScale mm}
\psset{linewidth=0.3,dotsep=1,hatchwidth=0.3,hatchsep=1.5,shadowsize=1,dimen=middle}
\psset{dotsize=0.7 2.5,dotscale=1 1,fillcolor=black}
\psset{arrowsize=1 2,arrowlength=1,arrowinset=0.25,tbarsize=0.7 5,bracketlength=0.15,rbracketlength=0.15}
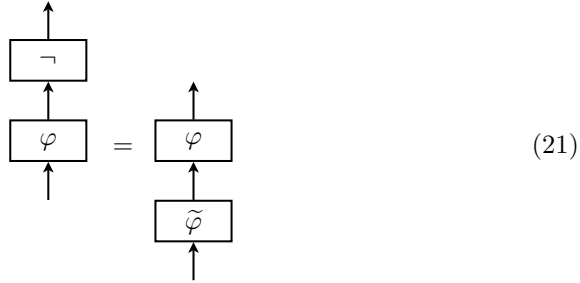
\begin{pspicture}(0,0)(10.62,22.5)
\psline{->}(5,-6.88)(5,-1.24)
\pspolygon[](-0.62,4.88)(10.62,4.88)(10.62,-1.12)(-0.62,-1.12)
\psline{->}(5,4.99)(5,10.62)
\rput(5,1.76){$\ahh$}
\pspolygon[](-0.62,16.76)(10.62,16.76)(10.62,10.75)(-0.62,10.75)
\psline{->}(5,16.88)(5,22.5)
\rput(5,13.75){$\chh$}
\end{pspicture}
 & = & \def\JPicScale{.9}\newcommand{\ahh}{\widetilde \varphi}\newcommand{\chh}{\varphi}  
\ifx\JPicScale\undefined\def\JPicScale{1}\fi
\psset{unit=\JPicScale mm}
\psset{linewidth=0.3,dotsep=1,hatchwidth=0.3,hatchsep=1.5,shadowsize=1,dimen=middle}
\psset{dotsize=0.7 2.5,dotscale=1 1,fillcolor=black}
\psset{arrowsize=1 2,arrowlength=1,arrowinset=0.25,tbarsize=0.7 5,bracketlength=0.15,rbracketlength=0.15}
\begin{pspicture}(0,0)(10.62,10.62)
\psline{->}(5,-18.75)(5,-13.12)
\pspolygon[](-0.62,-7)(10.62,-7)(10.62,-13)(-0.62,-13)
\psline{->}(5,-6.88)(5,-1.25)
\rput(5,-10.12){$\ahh$}
\pspolygon[](-0.62,4.88)(10.62,4.88)(10.62,-1.12)(-0.62,-1.12)
\psline{->}(5,5)(5,10.62)
\rput(5,1.88){$\chh$}
\end{pspicture}

\eea

\vspace{3\baselineskip}
\ee{defn}

\paragraph{Explanation.} Recall from Eq.~\eqref{eq-logic} that the computation $\neg : \Nn\to \Nn$ implements the logical negation, i.e. $\neg \circ \tru = \fls$ and $\neg\circ \fls = \tru$. The informal idea behind the computation $\widetilde \varphi:L\to L$ is that it maps the elements of the set $\{x|\varphi(x)\}$ into $\{x|\neg \varphi(x)\}$, and \emph{vice versa}, thus leading to $\varphi\big(\widetilde \varphi(x)\big) \iff \neg \varphi(x)$. This is the intuition behind Eq.~\eqref{eq-tilde}. If we work with sets, then such a  $\widetilde \varphi$ switch of $\{x|\varphi(x)\}$ and $\{x|\neg \varphi(x)\}$  is possible whenever both sets are nonempy, i.e. whenever the predicate $\varphi(x)$ is neither always true, not always false. This is why we call the predicates $\varphi$ which allow $\widetilde \varphi$ \emph{nontrivial}.

\be{lemma}\label{lemma-nontriv}
The predicate $\downarrow$, defined in Eq.~\eqref{eq-charnum}, is nontrivial.
\ee{lemma}

\bprf{ of Lemma~\ref{lemma-nontriv}}
The computation $\widetilde\downarrow :\Nn\to \Nn$, which we simply write as $\thicksim $ can be defined by
\bear
\thicksim  & = & d\, \circ \downarrow \mbox{ with}\\
\enco d  & = & \lambda x.\  \ifthenelse\left({\sf iszero}\left(\rho (x)\right)\right) \bot \enco 1
\eear
where $\rho$ computes the predecessor, and maps 0 to itself, whereas ${\sf iszero}$ maps $0$ to $1$ and the other numbers to $0$. The result is that $d$ maps 1 to $\bot$ and $0$ to 1, so that $\thicksim $ maps 
\begin{itemize}
\item $\bot \stackrel\downarrow \longmapsto 0 \stackrel d  \longmapsto 1$ and 
\item $n \stackrel\downarrow \longmapsto 1 \stackrel d  \longmapsto \bot$, for all $n\in \NNn$.
\end{itemize}
\epr

\bprf{ of Prop.~\ref{prop-halting}}
If $H$ is decidable then
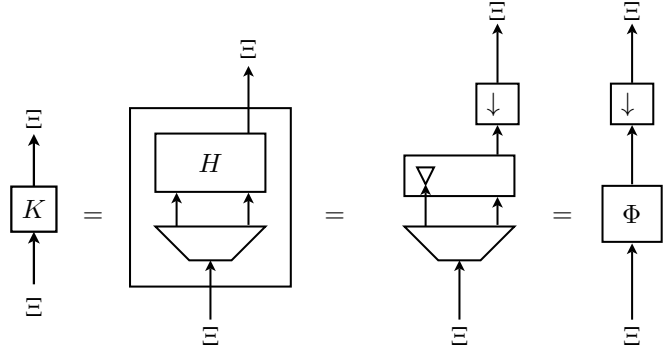
\bear
\def\JPicScale{1}\newcommand{\aah}{\Nn}\renewcommand{\dh}{\Nn}\newcommand{\ahh}{K} 
\ifx\JPicScale\undefined\def\JPicScale{1}\fi
\psset{unit=\JPicScale mm}
\psset{linewidth=0.3,dotsep=1,hatchwidth=0.3,hatchsep=1.5,shadowsize=1,dimen=middle}
\psset{dotsize=0.7 2.5,dotscale=1 1,fillcolor=black}
\psset{arrowsize=1 2,arrowlength=1,arrowinset=0.25,tbarsize=0.7 5,bracketlength=0.15,rbracketlength=0.15}
\begin{pspicture}(0,0)(8,13.75)
\rput(5,13.75){$\dh$}
\psline{->}(5,-8.25)(5,-1.25)
\pspolygon[](2,4.75)(8,4.75)(8,-1.25)(2,-1.25)
\psline{->}(5,4.75)(5,11.75)
\rput(5,1.75){$\ahh$}
\rput(5,-11.25){$\aah$}
\end{pspicture}
 & = & \def\JPicScale{.9}\newcommand{\aah}{\Phi}\renewcommand{\dh}{\Nn}\newcommand{\ahh}{H} \newcommand{\diagonal}{}\newcommand{\charct}{\downarrow}
\ifx\JPicScale\undefined\def\JPicScale{1}\fi
\psset{unit=\JPicScale mm}
\psset{linewidth=0.3,dotsep=1,hatchwidth=0.3,hatchsep=1.5,shadowsize=1,dimen=middle}
\psset{dotsize=0.7 2.5,dotscale=1 1,fillcolor=black}
\psset{arrowsize=1 2,arrowlength=1,arrowinset=0.25,tbarsize=0.7 5,bracketlength=0.15,rbracketlength=0.15}
\begin{pspicture}(0,0)(23.75,25.62)
\rput(17.5,25.62){$\dh$}
\psline{->}(6.88,-0.62)(6.88,4.38)
\pspolygon[](3.75,13.12)(20,13.12)(20,4.5)(3.75,4.5)
\psline{->}(17.5,13.12)(17.5,23.12)
\rput(11.88,8.75){$\ahh$}
\psline{->}(17.5,-0.38)(17.5,4.38)
\pspolygon(3.75,-0.62)
(20,-0.62)
(15,-5.62)
(8.75,-5.62)(3.75,-0.62)
\psline{->}(11.88,-14.38)(11.88,-5.62)
\rput(11.88,-3.12){$\diagonal$}
\pspolygon[](0,16.88)(23.75,16.88)(23.75,-9.5)(0,-9.5)
\rput(11.88,-16.88){$\dh$}
\end{pspicture}
 \ \ \ = \ \ \ \renewcommand{\ahh}{}
\ifx\JPicScale\undefined\def\JPicScale{1}\fi
\psset{unit=\JPicScale mm}
\psset{linewidth=0.3,dotsep=1,hatchwidth=0.3,hatchsep=1.5,shadowsize=1,dimen=middle}
\psset{dotsize=0.7 2.5,dotscale=1 1,fillcolor=black}
\psset{arrowsize=1 2,arrowlength=1,arrowinset=0.25,tbarsize=0.7 5,bracketlength=0.15,rbracketlength=0.15}
\begin{pspicture}(0,0)(20.62,31.25)
\rput(17.5,31.25){$\dh$}
\psline{->}(6.88,-0.62)(6.88,5.62)
\psline{->}(17.5,20.62)(17.5,29.38)
\psline{->}(17.5,-0.38)(17.5,3.75)
\pspolygon(3.75,-0.62)
(20,-0.62)
(15,-5.62)
(8.75,-5.62)(3.75,-0.62)
\psline{->}(11.88,-14.38)(11.88,-5.62)
\rput(11.88,-3.12){$\diagonal$}
\rput(11.88,-16.88){$\dh$}
\pspolygon[](3.75,9.88)(20,9.88)(20,3.88)(3.75,3.88)
\rput(16.88,6.88){$\ahh$}
\psline{->}(17.5,10)(17.5,14.38)
\pspolygon[](14.38,20.5)(20.62,20.5)(20.62,14.5)(14.38,14.5)
\rput(16.88,17.5){$\charct$}
\pspolygon(5.62,8.12)
(8.12,8.12)
(6.88,5.62)(5.62,8.12)
\end{pspicture}
 \ \ \ =\ \ \ \renewcommand{\ahh}{\Phi}
\ifx\JPicScale\undefined\def\JPicScale{1}\fi
\psset{unit=\JPicScale mm}
\psset{linewidth=0.3,dotsep=1,hatchwidth=0.3,hatchsep=1.5,shadowsize=1,dimen=middle}
\psset{dotsize=0.7 2.5,dotscale=1 1,fillcolor=black}
\psset{arrowsize=1 2,arrowlength=1,arrowinset=0.25,tbarsize=0.7 5,bracketlength=0.15,rbracketlength=0.15}
\begin{pspicture}(0,0)(8.12,31.25)
\rput(3.75,31.25){$\dh$}
\psline{->}(3.75,20.62)(3.75,29.38)
\psline{->}(3.75,-14.39)(3.75,-3.12)
\rput(3.75,-16.88){$\dh$}
\pspolygon[](-0.62,5.39)(8.12,5.39)(8.12,-2.89)(-0.62,-2.89)
\rput(3.75,1.25){$\ahh$}
\psline{->}(3.75,5.62)(3.75,14.38)
\pspolygon[](0.62,20.5)(6.88,20.5)(6.88,14.5)(0.62,14.5)
\rput(3.12,17.5){$\charct$}
\end{pspicture}

\eear

\vspace{3\baselineskip}
is decidable too. But from Lemma~\ref{lemma-nontriv} we have a computation $\thicksim $ such that $\downarrow \circ \thicksim\  = \neg\, \circ \downarrow$, and hence
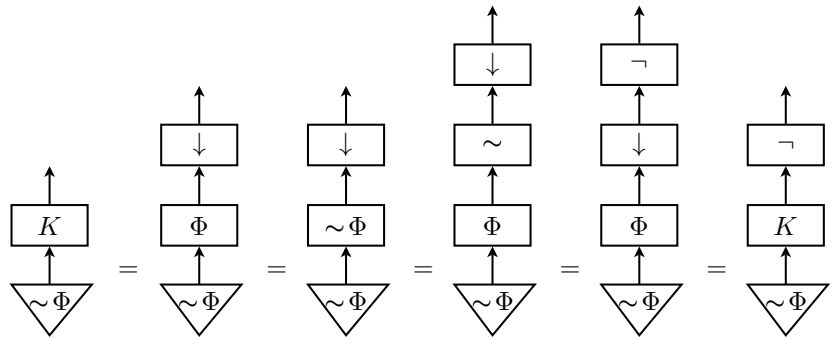
\bear
\def\JPicScale{.9}\newcommand{\ahh}{K}\newcommand{\bhh}{\thicksim\!\Phi}  
\ifx\JPicScale\undefined\def\JPicScale{1}\fi
\psset{unit=\JPicScale mm}
\psset{linewidth=0.3,dotsep=1,hatchwidth=0.3,hatchsep=1.5,shadowsize=1,dimen=middle}
\psset{dotsize=0.7 2.5,dotscale=1 1,fillcolor=black}
\psset{arrowsize=1 2,arrowlength=1,arrowinset=0.25,tbarsize=0.7 5,bracketlength=0.15,rbracketlength=0.15}
\begin{pspicture}(0,0)(11.25,16.25)
\psline{->}(5,-1.25)(5,4.38)
\pspolygon[](-0.62,10.5)(10.62,10.5)(10.62,4.5)(-0.62,4.5)
\psline{->}(5,10.62)(5,16.25)
\rput(5,7.38){$\ahh$}
\rput(5,-3.75){$\bhh$}
\pspolygon(-0.62,-1.25)
(11.25,-1.25)
(5,-8.75)(-0.62,-1.25)
\end{pspicture}
 & = & \def\JPicScale{.9}\newcommand{\ahh}{\Phi}\newcommand{\bhh}{\thicksim\!\Phi}\newcommand{\chh}{\downarrow}  \ \  =\ \  \def\JPicScale{.9}\renewcommand{\ahh}{\thicksim\! \Phi}  \ \  =\ \  \def\JPicScale{.9}\renewcommand{\ahh}{\Phi}\renewcommand{\chh}{\thicksim} \newcommand{\dhh}{\downarrow}  
\ifx\JPicScale\undefined\def\JPicScale{1}\fi
\psset{unit=\JPicScale mm}
\psset{linewidth=0.3,dotsep=1,hatchwidth=0.3,hatchsep=1.5,shadowsize=1,dimen=middle}
\psset{dotsize=0.7 2.5,dotscale=1 1,fillcolor=black}
\psset{arrowsize=1 2,arrowlength=1,arrowinset=0.25,tbarsize=0.7 5,bracketlength=0.15,rbracketlength=0.15}
\begin{pspicture}(0,0)(11.25,40)
\psline{->}(5,-1.25)(5,4.38)
\pspolygon[](-0.62,10.5)(10.62,10.5)(10.62,4.5)(-0.62,4.5)
\psline{->}(5,10.62)(5,16.25)
\rput(5,7.38){$\ahh$}
\rput(5,-3.75){$\bhh$}
\pspolygon(-0.62,-1.25)
(11.25,-1.25)
(5,-8.75)(-0.62,-1.25)
\pspolygon[](-0.62,22.38)(10.62,22.38)(10.62,16.38)(-0.62,16.38)
\psline{->}(5,22.5)(5,28.12)
\rput(5,19.38){$\chh$}
\pspolygon[](-0.62,34.25)(10.62,34.25)(10.62,28.26)(-0.62,28.26)
\psline{->}(5,34.38)(5,40)
\rput(5,31.25){$\dhh$}
\end{pspicture}
\ \  =\ \  \def\JPicScale{.9}\renewcommand{\ahh}{\Phi}\renewcommand{\chh}{\downarrow} \renewcommand{\dhh}{\neg}  \ \  =\ \  \def\JPicScale{.9}\renewcommand{\ahh}{K} \renewcommand{\chh}{\neg}  
\eear

\vspace{1.5\baselineskip}
where we write $\thicksim\!\Phi$ instead of $\Phi\comp \thicksim$. But $K(\thicksim\! \Phi) = \neg K(\thicksim\! \Phi)$ implies that $K(\thicksim\! \Phi) \not\in \Bits$. So $K$ cannot be decidable, and thus $H$ is not decidable either.
\epr

\section{Rice's Theorem}\label{Sec-Rice}
\be{defn}\label{def-ext-pred}
We say that a predicate $\alpha : \Nn\to \Nn$ is \emph{over computations} if for all $p,q\in \Nn$
\bear
\{p\} = \{q\} & \Longrightarrow & \alpha \circ p = \alpha \circ q
\eear
\ee{defn}

\be{prop}\label{prop-rice}
Every nontrivial predicate over computations is undecidable.
\ee{prop}

\bpr
Let $\alpha:\Nn\to \Nn$ be a nontrivial extensional predicate. Since it is nontrivial, there is a computation $\thicksim\ =\ \widetilde \alpha: \Nn\to\Nn$ such that $\neg\, \circ \alpha = \alpha\, \circ \thicksim$.  Define the computation

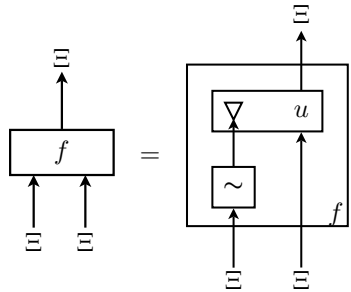
\bear
\def\JPicScale{1}\newcommand{\eh}{\Nn}\newcommand{\fh}{\Nn}\renewcommand{\dh}{\Nn}\newcommand{\ahh}{f}  & = & \def\JPicScale{.9}\newcommand{\aah}{f}\renewcommand{\dh}{\Nn}\newcommand{\ahh}{\UK} \newcommand{\diagonal}{\thicksim}\newcommand{\eh}{\Nn}\newcommand{\fh}{\Nn}
\eear

\vspace{3\baselineskip}

By Corollary~\ref{corollary-kleene}, $f$ has a fixed program $p$, which means
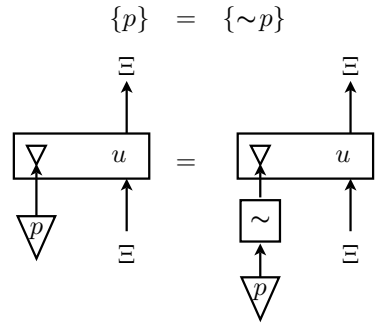
\bear \{p\} & = & \{\thicksim\! p\}\\[2ex]
\def\JPicScale{1}\newcommand{\aah}{p}\renewcommand{\dh}{\Nn}\newcommand{\ahh}{\UK}\newcommand{\bhh}{\Nn}  & = & \def\JPicScale{1}\newcommand{\aah}{p}\renewcommand{\dh}{\Nn}\newcommand{\ahh}{\UK}\newcommand{\bhh}{\Nn} \newcommand{\chh}{\thicksim} 
\ifx\JPicScale\undefined\def\JPicScale{1}\fi
\psset{unit=\JPicScale mm}
\psset{linewidth=0.3,dotsep=1,hatchwidth=0.3,hatchsep=1.5,shadowsize=1,dimen=middle}
\psset{dotsize=0.7 2.5,dotscale=1 1,fillcolor=black}
\psset{arrowsize=1 2,arrowlength=1,arrowinset=0.25,tbarsize=0.7 5,bracketlength=0.15,rbracketlength=0.15}
\begin{pspicture}(0,0)(20,13.75)
\rput(17,13.75){$\dh$}
\psline{->}(5,-3.75)(5,0.75)
\pspolygon[](2,4.75)(20,4.75)(20,-1.25)(2,-1.25)
\psline{->}(17,4.75)(17,11.75)
\rput(16,1.75){$\ahh$}
\rput(5,-16.25){$\aah$}
\psline{->}(17,-8.25)(17,-1.25)
\rput(17,-11.25){$\bhh$}
\pspolygon[](2.38,-4.12)(7.62,-4.12)(7.62,-9.62)(2.38,-9.62)
\psline{->}(5,-14.38)(5,-9.88)
\rput(5,-6.88){$\chh$}
\pspolygon(3.75,3.12)
(6.25,3.12)
(5,0.62)(3.75,3.12)
\pspolygon(2.5,-14.38)
(7.5,-14.38)
(5.01,-20)(2.5,-14.38)
\end{pspicture}

\eear 

\vspace{4\baselineskip}

But since $\alpha$ is by assumption over computations, 
\bear
\{p\} = \{\thicksim\! p\} & \Longrightarrow & \alpha \circ p = \alpha\, \circ \thicksim\! \circ\, p
\eear
and thus
\newpage
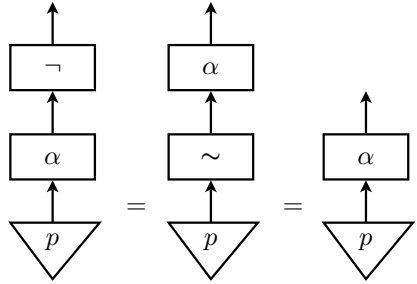
\bear
\def\JPicScale{1}\newcommand{\ahh}{\alpha}\newcommand{\bhh}{p}\newcommand{\chh}{\neg}   & = & \def\JPicScale{1}\newcommand{\ahh}{\thicksim}\newcommand{\bhh}{p}\newcommand{\chh}{\alpha}  \ \ =\ \ 
\def\JPicScale{1}\renewcommand{\ahh}{\alpha}\renewcommand{\bhh}{p}  
\ifx\JPicScale\undefined\def\JPicScale{1}\fi
\psset{unit=\JPicScale mm}
\psset{linewidth=0.3,dotsep=1,hatchwidth=0.3,hatchsep=1.5,shadowsize=1,dimen=middle}
\psset{dotsize=0.7 2.5,dotscale=1 1,fillcolor=black}
\psset{arrowsize=1 2,arrowlength=1,arrowinset=0.25,tbarsize=0.7 5,bracketlength=0.15,rbracketlength=0.15}
\begin{pspicture}(0,0)(11.25,16.25)
\psline{->}(5,-1.25)(5,4.38)
\pspolygon[](-0.62,10.5)(10.62,10.5)(10.62,4.5)(-0.62,4.5)
\psline{->}(5,10.62)(5,16.25)
\rput(5,7.38){$\ahh$}
\rput(5,-3.75){$\bhh$}
\pspolygon(-0.62,-1.25)
(11.25,-1.25)
(5,-8.75)(-0.62,-1.25)
\end{pspicture}
 
\eear

\epr

\section{Future work}
Here we sketch the further technical developments of monoidal computer, that will be presented in the sequel papers.

\paragraph{Computable types.} Although not strictly necessary, it is often convenient to have a type $\{x\in L\ |\ \alpha(x)\}$ for every computable predicate $\alpha : L\to \Nn$ in a monoidal computer $\CCC$. Formally, this amounts to requiring that $\CCC$ has equalizers. This is what we shall call \emph{comprehensive\/} monoidal computer. In the concrete examples, this means that the types do not just represent the arities of the computations any more, i.e. the numbers of their inputs and outputs, but that every recursively enumerable set is now represented by a type. In particular, a comprehensive monoidal computer will thus have internal representations of $\Bits$ and $\NNn$ from Sec.~\ref{sec-numbers}, as well as their powers.

\paragraph{Program complexity.} As computations, the program complexity measures can be derived from the universal evaluators. We follow this idea in representing them in the monoidal computer. First of all, any program complexity measure $\CX^M_L\in \CCC(\Nn\otimes L,\Nn)$ should be defined on a program $p$ and a value $x$ if and only if $\UK_L^M(p,x) = \deco p(x)$ is defined, or more generally
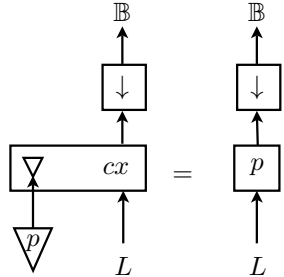
\bear
\def\JPicScale{1}\newcommand{\aah}{p}\newcommand{\bbh}{\downarrow}\renewcommand{\dh}{\Bits}\newcommand{\bhh}{L}\newcommand{\ahh}{\CX} 
\ifx\JPicScale\undefined\def\JPicScale{1}\fi
\psset{unit=\JPicScale mm}
\psset{linewidth=0.3,dotsep=1,hatchwidth=0.3,hatchsep=1.5,shadowsize=1,dimen=middle}
\psset{dotsize=0.7 2.5,dotscale=1 1,fillcolor=black}
\psset{arrowsize=1 2,arrowlength=1,arrowinset=0.25,tbarsize=0.7 5,bracketlength=0.15,rbracketlength=0.15}
\begin{pspicture}(0,0)(20,22.5)
\psline{->}(5,-6.25)(5,0.75)
\pspolygon[](2,4.75)(20,4.75)(20,-1.25)(2,-1.25)
\rput(16,1.75){$\ahh$}
\rput(5,-8.12){$\aah$}
\psline{->}(17,-8.25)(17,-1.25)
\rput(17,-11.25){$\bhh$}
\pspolygon(2.5,-6.25)
(7.5,-6.25)
(5.01,-11.88)(2.5,-6.25)
\pspolygon(3.75,3.12)
(6.25,3.12)
(5,0.62)(3.75,3.12)
\rput(16.88,22.5){$\dh$}
\psline{->}(16.88,5)(16.88,9.38)
\pspolygon[](14.25,15.5)(20,15.5)(20,9.5)(14.25,9.5)
\psline{->}(16.88,15.62)(16.88,20.62)
\rput(16.88,12.5){$\bbh$}
\end{pspicture}
 & = & \def\JPicScale{1}\newcommand{\aah}{p}\renewcommand{\dh}{\Bits}\newcommand{\ahh}{f} \newcommand{\bhh}{L} \newcommand{\bbh}{\downarrow} 
\ifx\JPicScale\undefined\def\JPicScale{1}\fi
\psset{unit=\JPicScale mm}
\psset{linewidth=0.3,dotsep=1,hatchwidth=0.3,hatchsep=1.5,shadowsize=1,dimen=middle}
\psset{dotsize=0.7 2.5,dotscale=1 1,fillcolor=black}
\psset{arrowsize=1 2,arrowlength=1,arrowinset=0.25,tbarsize=0.7 5,bracketlength=0.15,rbracketlength=0.15}
\begin{pspicture}(0,0)(8.12,22.5)
\psline{->}(5,-8.25)(5,-1.25)
\pspolygon[](2,4.75)(8,4.75)(8,-1.25)(2,-1.25)
\rput(5,1.88){$\aah$}
\rput(5,22.5){$\dh$}
\psline{->}(5.12,4.75)(5,9.38)
\pspolygon[](2.38,15.5)(8.12,15.5)(8.12,9.5)(2.38,9.5)
\psline{->}(5,15.62)(5,20.62)
\rput(5.12,-11.25){$\bhh$}
\rput(5,12.5){$\bbh$}
\end{pspicture}
 
\eear

\vspace{2.3\baselineskip}
where $\downarrow$ is the predicate from Def.~\ref{def-numeric}. The main requirement is that all programs are composed at with at most a constant cost in complexity, i.e.
\bear
\CX^N_L\left((p\comp q), x\right) & \leqad& \CX_L^M(p, x) + \CX_M^N(q,px)\\
\CX_{L_1\otimes L_2}^{M_1\otimes M_2}(p_1\otimes p_2, x_1\otimes x_2) &\leqad & \CX_{L_1}(p_1,x_1) + \CX_{L_2}(p_2,x_2)
\eear
where $\leqad$ is the "upto" order of functions
\bear
f\leqad g &\iff & \exists c\in \NNn \ \forall n\in \NNn.\ fn \leq c+gn
\eear
Finally, we also need a \emph{length}\/ function $\ell \in \Base \CCC(\Nn,\NNn)$ with
\[\ell n \leqad n\qquad \mbox{ and } \qquad \ell (p\comp q) \eqad \ell p + \ell q \eqad \ell (p \otimes q)\]
such that every complexity measure satisfies
\bear
\CX_\Nn(\ell,x) & \eqad & \ell x
\eear 
A computation that implements the length function can be thought of as just reading its input, so that this last requirement just says that the needed space and time is just the length of the input, plus the constant length of the needed commands. The abstract versions of the time and the space complexity measures can be introduced by requiring that they satisfy 
\bear
\TIME_L^N\left((p\comp q), x\right) & \eqad & \TIME_L^M(p, x) + \TIME_M^N(q,px)\\
\TIME_{L_1\otimes L_2}^{M_1\otimes M_2}(p_1\otimes p_2, x_1\otimes x_2)& \eqad & \max\big\{\TIME_{L_1}(p_1,x_1), \TIME_{L_@}(p_2,x_2)\big\}
\eear
and
\bear
\SPACE_L^N\left((p\comp q), x\right) &  \eqad & \max\big\{\SPACE_L^M(p, x), \SPACE_M^N(q,px)\big\}\\
\SPACE_{L_1\otimes L_2}^{M_1\otimes M_2}(p_1\otimes p_2, x_1\otimes x_2)& \eqad & \SPACE_{L_1}(p_1,x_1)+ \SPACE_{L_@}(p_2,x_2)
\eear

\paragraph{Randomized computation.} For any set $X$, denote by $\DDD X$ the set of finitely supported subprobability distributions over $X$, i.e.
\bear
\DDD X & = & \left\{P:X\to [0,1]\ |\ \supp P\lt \infty \wedge \sum_{x\in X} Px \leq 1 \right\}
\eear
where $\supp P$ is the cardinality of the set $\{x\in X| Px\neq 0\}$. For any category $\CCC$ we can now define the \emph{randomized version}\/ $\RC\CCC$ by setting
\begin{itemize}
\item objects: $|\RC\CCC| \ = \ |\CCC|$
\item morphisms: $\RC\CCC (A,B) \ = \  \DDD\CCC(A,B)$
\item composition: 
\begin{gather*}
\prooftree
\Phi\ :\ \CCC(A,B)\to [0,1] \qquad \qquad \Psi\ :\ \CCC(B,C)\to [0,1]
\justifies
(\Psi\circ \Phi)\ :\ \CCC(A,C)\to [0,1]
\endprooftree\\[1ex]
(\Psi\circ \Phi)_h\ \ =\ \ \sum_{g\circ f = h} \Psi_g \cdot \Phi_f
\end{gather*}
\item identities: $\iota_A:\CCC(A,A)\to [0,1]$ is $\iota_A(f) = 1$ if and only if $f=\id_A$, otherwise 0.
\end{itemize}
The monoidal structure, the data services, and the induced operations have  natural liftings from $\CCC$ to $\RC\CCC$. But what is a randomized monoidal computer? How do the universal evaluators interpret randomized programs?

In practice, randomized programs are implemented as ordinary programs which, in addition to their normal inputs,  also input an additional argument, where they receive random seeds. More precisely, given an ordinary computation $f\in \CCC(S\otimes L,M)$, where $S\in |\CCC|$ is taken to be the type of random seeds, and a distribution $\varsigma \in \DDD\CCC(S)$ we can define
\begin{gather*}
\prooftree
f \in \CCC(S\otimes L, M)\qquad \qquad \varsigma\in \DDD\CCC(S)
\justifies
f^\varsigma \in \DDD \CCC(L,M)
\endprooftree\\[2ex]
f^\varsigma_g\  = \ \Pr(g \stackrel \varsigma \leftarrow f) \ = \ \sum_{s_{SL}(f,r) = g} \varsigma r
\end{gather*}
Formally, the random seeds of type $S$ can be denoted by an indeterminate element $x$ of type $S$ in the polynomial category $\CCC[x:S]$ \cite{Lambek-Scott:book,PavlovicD:MSCS97,PavlovicD:Qabs}. Indeed, we randomize a value, e.g. in a security protocol, when we need to assure that no one can predict it, or derive it from any other values; and an indeterminate element $x$ satisfies similar requirements: it cannot be algebraically derived from any other element, and it is equally likely to denote any of them.     Assigning a distribution $\varsigma \in \DDD\CCC(S)$ to a random variable $x:S$ adjointed to $\CCC$ induces a functor $\CCC[x:S]\tto\varsigma \DDD\CCC$, which summarizes the  above derivation of  a randomized computation $f^\varsigma\in \DDD \CCC(L,M)$ from a seeded computation $f\in \CCC(S\otimes L, M)$.  When $\CCC(S)$ is finite, say $S= \{0,1\}^n$, and when $\varsigma$ is the uniform distribution, then the above definition of $f^\varsigma$ boils down to the usual view of the input-output probability
\bear
\Pr\left(b \stackrel{\$} \leftarrow fa\right) & = & \frac{\#\{r\in \{0,1\}^n\ |\ f(r,a) = b\}}{2^n}
\eear
for $a\in \CCC(L)$ and $b\in \CCC(M)$. 

However, to capture the general computations, admitting the inputs of varied lengths, which usually also requires seeds of varied lengths, this simple idea needs to be extended to \emph{ensembles} of computations, which furthermore need to be taken modulo computational \emph{indistinguishability} \cite{GoldreichO:book-1}. This is where the convolution will play a pivotal role.

\bibliography{ref-resource,PavlovicD}
\bibliographystyle{plain}

\end{document}